\documentclass[twocolumn,pra]{revtex4}
\usepackage[latin9]{inputenc}
\usepackage{xcolor}
\usepackage{textcomp}
\usepackage{amsmath}
\usepackage{amssymb}
\usepackage{graphicx}

\makeatletter
\@ifundefined{textcolor}{}
{%
 \definecolor{BLACK}{gray}{0}
 \definecolor{WHITE}{gray}{1}
 \definecolor{RED}{rgb}{1,0,0}
 \definecolor{GREEN}{rgb}{0,1,0}
 \definecolor{BLUE}{rgb}{0,0,1}
 \definecolor{CYAN}{cmyk}{1,0,0,0}
 \definecolor{MAGENTA}{cmyk}{0,1,0,0}
 \definecolor{YELLOW}{cmyk}{0,0,1,0}
}

\usepackage{amsfonts}

\makeatother

\begin{document}
\title{Nonclassical phonon pair}
\author{Yu Wang$^{1,2,3\#}$, Zhen Shen$^{1,2,3\#}$, Mai Zhang$^{1,2,3\#}$,
Zhi-Peng Shi$^{1,2,3\#}$, Hong-Yi Kuang$^{1,2,3}$, Shuai Wan$^{1,2,3}$,
Fang-Wen Sun$^{1,2,3}$, Guang-Can Guo$^{1,2,3}$, and Chun-Hua Dong$^{1,2,3\text{,*}}$}
\affiliation{$^{1}$CAS Key Laboratory of Quantum Information, University of Science
and Technology of China, Hefei, Anhui 230026, China}
\affiliation{$^{2}$CAS Center For Excellence in Quantum Information and Quantum
Physics, University of Science and Technology of China, Hefei, Anhui
230088, China}
\affiliation{$^{3}$Hefei National Laboratory, University of Science and Technology
of China, Hefei, Anhui 230088, China}
\email{chunhua@ustc.edu.cn}

\date{\today}
\begin{abstract}
Quantum-correlated photon pairs are crucial resources for modern quantum
information science. Similarly, the reliable generation of nonclassical
phonon pairs is vital for advancing engineerable solid-state quantum
devices and hybrid quantum networks based on phonons. Here, we present
a novel approach to\textbf{ }generate quantum-correlated phonon pairs
in a suspended silicon microstructure initialized in its motional
ground state. By simultaneously implementing red- and blue-detuned
laser pulses, equivalent high-order optomechanical nonlinearity\textendash \textendash specifically,
an effective optomechanical four-wave mixing process\textendash \textendash is
achieved for generating a nonclassical phonon pair, which is then
read out via a subsequent red-detuned pulse. We demonstrate the nonclassical
nature of the generated phonon pair through the violation of the Cauchy\textendash Schwarz
inequality. Our experimentally observed phonon pair violates the classical
bound by more than 5 standard deviations and maintains a decoherence
time of $132\:\mathrm{ns}$. This work reveals novel quantum manipulation
of phonon states enabled by equivalent high-order optomechanical nonlinearity
within a pulse scheme and provides a valuable quantum resource for\textbf{
}mechanical quantum computing.
\end{abstract}
\maketitle
\textbf{\emph{Introduction. -}}\emph{ }Quantum-correlated pairs are
crucial resources for modern quantum information science and underpin
a broad range of fundamental physical effects, including in distributed
quantum computation, quantum communication, and quantum metrology
\cite{kimble2008quantum,walther2005experimental,giovannetti2004quantum,o2009photonic,pezze2018quantum}.
Over the past decade, nonclassical particle pairs, particularly entangled
photon pairs, have played a pivotal role in demonstrating the triumph
of quantum physics over local realism through Bell inequality violations
and are key components in modern quantum information protocols such
as quantum repeaters, quantum memories, and device-independent quantum
key distribution \cite{giustina2015significant,sangouard2011quantum,liu2021heralded,wallucks2020quantum,acin2007device}.
Entangled pairs have been demonstrated not only between various encoding
photons \cite{pan2012multiphoton,sheremet2023waveguide,meesala2023non}
but also between photons and matter qubits, such as atoms \cite{volz2006observation},
spins \cite{gao2012observation}, phonons\cite{riedinger2016non},
and electrons \cite{feist2022cavity}. These unique quantum resources
are paramount for harnessing quantum correlations in hybrid quantum
networks \cite{clerk2020hybrid,shen2022coherent}.

Phonons, which are strikingly similar to photons, are emerging as
promising candidates for engineerable solid-state quantum devices
and quantum communication interfaces because of their ability to coherently
interact with various quantum systems \cite{aspelmeyer2014cavity,riedinger2016non,dong2012optomechanical,shen2022coherent,wang2024optomechanical}.
This has spurred significant interest in quantum manipulation of phonon
states, offering a novel platform in quantum information science.
Advances in opto- and electromechanical domains include ground-state
cooling \cite{chan2011laser}, Fock state phonon manipulation \cite{riedinger2016non,chu2018creation},
entanglement of mechanical oscillators \cite{riedinger2018remote,wollack2022quantum},
and quantum transducers \cite{meesala2023non}. In particular, significant
recent advancements in surface acoustic wave platforms driven by strongly
coupled superconducting circuits have enabled impressive demonstrations
of nonclassical phonon states, including multiphonon Fock states \cite{chu2018creation},
phononic beam splitters \cite{qiao2023splitting}, and mechanical
qubits \cite{yang2024mechanical}. For typical optomechanical devices,
which are key components of mechanical quantum elements, strong confinement
of phonons leads to a longer lifetime and relatively small footprints,
making them well suited for scaling to more complex circuits and performing
multistage manipulation\cite{zivari2023single}. However, the generation
of nonclassical phonon pairs via optical control remains challenging
because of the limited strength of single photon optomechanical coupling
and the underdeveloped exploitation of equivalent high-order optomechanical
nonlinearity \cite{arrangoiz2019resolving,kuang2023nonlinear,yang2024mechanical}.

Here, we present a method for generating quantum-correlated phonon
pairs from an optomechanical crystal (OMC) nanobeam cavity by leveraging
effective high-order optomechanical nonlinearity, in which Stokes
and anti-Stokes interactions are simultaneously implemented in one
pulse. Analogous to the spontaneous parametric down-conversion (SPDC)
process, the effective optomechanical spontaneous four-wave mixing
(SFWM) process in our protocol creates a degenerate correlated phonon
pair. The generated quantum correlated phonon pair is then retrieved
by a second red-detuned pulse and verified by the well-known Cauchy\textendash Schwarz
inequality, which is a standard criterion for distinguishing classical
from quantum correlations. Here we demonstrate not only that engineering
the equivalent high-order optomechanical nonlinearity under a pulse
scheme is possible but also that our protocol for quantum manipulation
of phonon states could generate quantum resources for mechanical quantum
computing.

\begin{figure*}[t]
\centerline{\includegraphics[clip,width=0.9\linewidth]{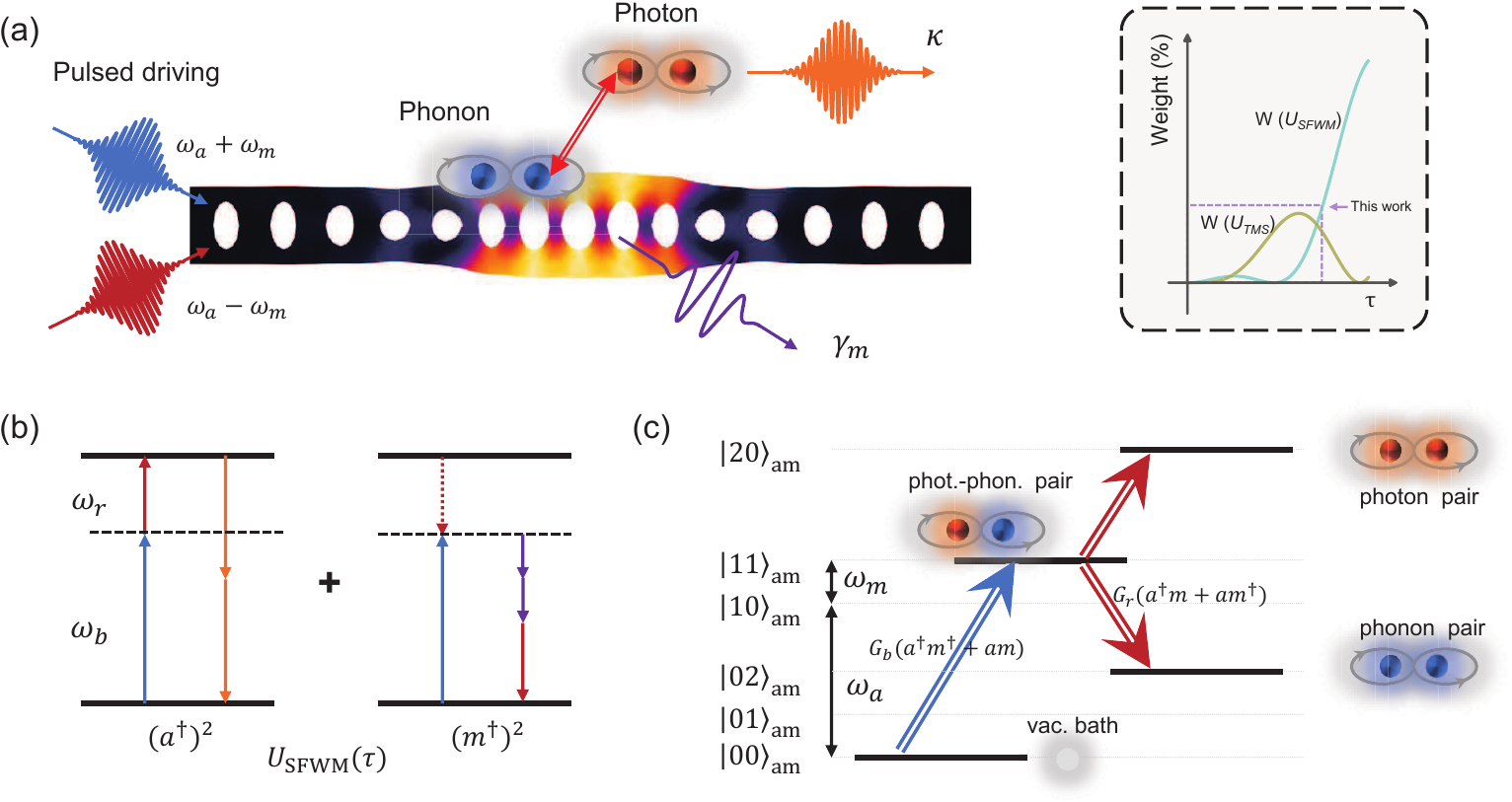}}\caption{\textbf{Schematic of the nonclassical phonon pair generation.} (a)
An OMC nanobeam cavity is simultaneously driven by optical pulses
at laser frequencies of $\omega_{\mathrm{a}}\pm\omega_{\mathrm{m}}$
to generate a photon pair $\left|20\right\rangle _{\mathrm{a,m}}$
or a phonon pair $\left|02\right\rangle _{\mathrm{a,m}}$. The inset
graphically illustrates the ``weight engineering'' of linear and
quadratic terms in $U(t)$ with $a^{\dagger}$ (or $m^{\dagger}$)
by adjusting the pulse duration $\tau$. (b) Energy level diagrams
of optomechanical SFWM for optical (left) and mechanical (right) effective
counterparts. The entire process corresponds to the equivalent high-order
optomechanical nonlinearity interaction in Eq. \ref{eq:1}. (c) Optomechanical
energy diagram describing a virtual process in which a photon\textendash phonon
pair $\left|11\right\rangle _{\mathrm{a,m}}$ is first generated by
a blue-detuned pump and then transitions to $\left|20\right\rangle _{\mathrm{a,m}}$
or $\left|02\right\rangle _{\mathrm{a,m}}$ via an interaction with
a red-detuned pump.}

\label{Fig1}
\end{figure*}

\textbf{\textsl{\emph{Schematic of the nonclassical phonon pair generation.}}}\emph{
-} Our protocol for generating quantum-correlated phonon pairs is
schematically illustrated in Fig. \ref{Fig1} (a). The optomechanical
cavity is simultaneously driven by optical pulses at laser frequencies
of $\omega_{\mathrm{a}}\pm\omega_{\mathrm{m}}$, where $\omega_{\mathrm{a(m)}}$
is the optical (mechanical) resonance. Two different types of interactions
on the basis of cavity-enhanced Stokes and anti-Stokes Raman scattering
occur at the same time, corresponding to ``two-mode squeezing''
and ``beam-splitter'' interactions \cite{aspelmeyer2014cavity}.
The time-evolution operator is derived as $U(t)=e^{-iH_{\mathrm{int}}t/\hbar}$,
where $H_{\mathrm{int}}/\hbar=g_{\mathrm{o}}\left(\alpha_{\mathrm{r}}a^{\dagger}m+\alpha_{\mathrm{b}}a^{\dagger}m^{\dagger}+H.C.\right)$.
Here $g_{\mathrm{o}}$ is the single photon optomechanical coupling
rate. $\alpha_{\mathrm{r(b)}}$ and $a$ ($m$) are the annihilation
operators for the red (blue)-detuned pump and the optical (mechanical)
mode, respectively. The general linear term $U_{\mathrm{TMS}}(t)$
in $U(t)$ describes \textquotedblleft two-mode squeezing\textquotedblright{}
on the basis of cavity-enhanced Stokes scattering \cite{aspelmeyer2014cavity}.
Unusually, the quadratic term in $U(t)$,
\begin{equation}
U_{\mathrm{SFWM}}\left(t\right)=c(t)\left[\alpha_{\mathrm{r}}\alpha_{\mathrm{b}}\left(a^{\dagger}\right)^{2}mm^{\dagger}+\alpha_{\mathrm{r}}^{\dagger}\alpha_{\mathrm{b}}aa^{\dagger}\left(m^{\dagger}\right)^{2}\right]+H.C.,\label{eq:1}
\end{equation}
represents an equivalent optomechanical SFWM process, where $c(t)$
is a time-dependent coefficient (see the Supplementary Information).
The whole energy diagram for this process, shown in Fig. \ref{Fig1}(b),
is separated into optical and mechanical counterparts. The optical
counterpart derived from the first term in Eq. (\ref{eq:1}), is analogous
to the optical $\chi^{\left(3\right)}$ SFWM process \cite{silverstone2014chip}.
The mechanical counterpart derived from the second term in Eq. (\ref{eq:1})
describes the generation of an entangled phonon pair at $\omega_{\mathrm{m}}$
via absorption of a pump photon $\alpha_{\mathrm{b}}$ and release
of a pump photon $\alpha_{\mathrm{r}}$. This can be equivalent to
mechanical SPDC \cite{pan2012multiphoton}, in which an entangled
phonon pair at $\omega_{\mathrm{m}}$ is generated by the annihilation
of a mechanical pump quantum at $2\omega_{\mathrm{m}}$.

\begin{figure*}[t]
\centerline{\includegraphics[clip,width=0.8\textwidth]{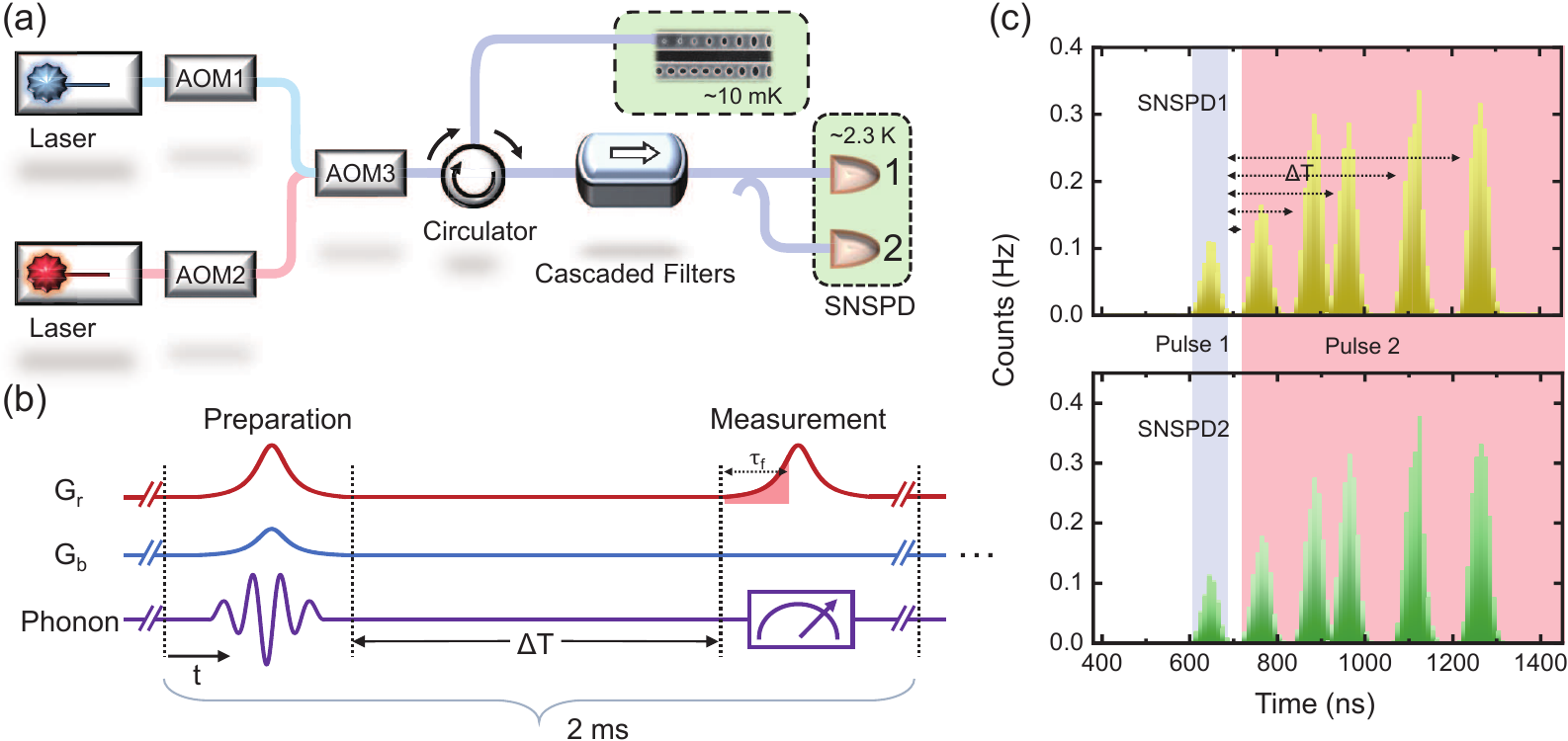}}\caption{(a) Experimental set-up. AOM, acousto-optic modulator; SNSPD, superconducting
nanowire single-photon detector. (b) Schematic of the preparation-measurement
protocol. The desired optomechanical quantum state $\left|\Phi\right\rangle _{\mathrm{a,m}}$
is prepared via the first pulses and measured by the second red-detuned
pulse which coherently retrieves phonons into signal photons and directly
checks the nonclassical correlation of the phonon pair. The red shadow
region with duration $\tau_{f}$ in the measurement pulse indicates
the duration used for the statistics of twofold coincidence events.
(c) Normalized counts of signal photons ($\omega_{\mathrm{a}}$) launched
into two SNSPDs in the HBT setup during pulse 1 and pulse 2 for various
preparation-measurement time delays $\Delta T$. The purple and pink
shadow areas represent the results for pulses 1 and 2, respectively
.}

\label{Fig2}
\end{figure*}
Therefore, when these optical pulses of duration $\tau$ are applied
to the initial optomechanical vacuum state $\left|00\right\rangle _{\mathrm{a,m}}$,
the system evolves from the initial vacuum state to $\left|\varphi\right\rangle _{\mathrm{a,m}}=U\left(\tau\right)\left|00\right\rangle _{\mathrm{a,m}}$.
The resulting optomechanical state is $\left|00\right\rangle _{\mathrm{a,m}}+\sqrt{p_{\mathrm{b}}(1-p_{\mathrm{r}})}\left|11\right\rangle _{\mathrm{a,m}}+\sqrt{p_{\mathrm{b}}p_{\mathrm{r}}}\left|\Phi\right\rangle _{\mathrm{a,m}}+O(p_{\mathrm{b}}),$
where
\begin{align}
\left|\Phi\right\rangle _{\mathrm{a,m}} & =\frac{1}{\sqrt{2}}\left(\left|20\right\rangle _{\mathrm{a,m}}+\left|02\right\rangle _{\mathrm{a,m}}\right)
\end{align}
is an optomechanical $NOON$ state \cite{afek2010high}, and $p_{\mathrm{r}}=\sin^{2}(2G_{\mathrm{r}}\tau)$
and $p_{\mathrm{b}}=G_{\mathrm{b}}^{2}\tau^{2}$ represent the mean
absorption probabilities of intracavity photons $n_{\mathrm{r(b)}}$,
with $G_{\mathrm{r(b)}}=g_{\mathrm{o}}\sqrt{n_{\mathrm{r(b)}}}$.
The evolution of $U_{\mathrm{SFWM}}(t)$ can be intuitively described
as a process in which a photon\textendash phonon pair is first generated
by absorption of a blue-detuned pump photon and subsequently transitions
to either photon pair $\left|20\right\rangle _{\mathrm{a,m}}$ or
phonon pair $\left|02\right\rangle _{\mathrm{a,m}}$ via absorption
or release of a red-detuned pump photon, as shown in Fig. \ref{Fig1}(c).
Specifically, when $p_{\mathrm{r}}=1$, a purer optomechanical $NOON$
state $\left|\Phi\right\rangle _{\mathrm{a,m}}$ can be created while
the photon\textendash phonon pair $\left|11\right\rangle _{\mathrm{a,m}}$
is suppressed, which is inaccessible in the previous continuous driving
quantum squeezing system \cite{wollman2015quantum}. Our pulse protocol
allows engineering of the weights of different order terms in $U(t)$
by controlling $G_{\mathrm{r}}$ and $\tau$, which promises the ability
to highlight equivalent high-order optomechanical nonlinear effects,
as depicted in the inset of Fig. \ref{Fig1}(a). Notably, the resulting
optomechanical states are distinct from the squeezing state and remain
largely unexplored (see Supplementary Information). Limited by $g_{\mathrm{o}}$
and laser heating, our experimental demonstration is performed under
$p_{\mathrm{r}}<1$, but this affects only the generatiion rates and
purity of $\left|\Phi\right\rangle _{\mathrm{a,m}}$.

\textbf{\textsl{\emph{Silicon optomechanical system. }}}\emph{-} The
experimental setup is shown in Fig. \ref{Fig2}(a). The OMC cavity,
fabricated from a silicon-on-insulator (SOI) wafer \cite{chan2012optimized,wang2023realization},
exhibits an optical cavity resonance at wavelength $\lambda_{\mathrm{a}}=1550.589\,\mathrm{nm}$
with a total damping rate $\kappa/2\pi\thickapprox700\,\mathrm{MHz}$
and a mechanical ``breathing'' mode at a frequency of $\omega_{\mathrm{m}}/2\pi\thickapprox5.2\,\mathrm{GHz}$
with a damping rate of $\gamma_{\mathrm{m}}/2\pi\thickapprox109\,\mathrm{kHz}$
in a dilution refrigerator. The optical and mechanical modes are coupled
via a combination of radiation pressure and photostriction with a
single-photon optomechanical coupling rate of $g_{\mathrm{o}}/2\pi=800\,\mathrm{kHz}$,
which is calibrated via sideband thermometry measurements \cite{riedinger2016non}.
In subsequent experiments, the pulse scheme, generated using acousto-optic
modulators (AOMs) to avoid laser heating and optomechanical backaction,
initializes the mechanical motion in its quantum ground state with
a mean thermal phonon occupation $n_{\mathrm{th}}=0.014\pm0.004$
(see the Supplementary Information).

\begin{figure*}[t]
\centerline{\includegraphics[clip,width=0.8\textwidth]{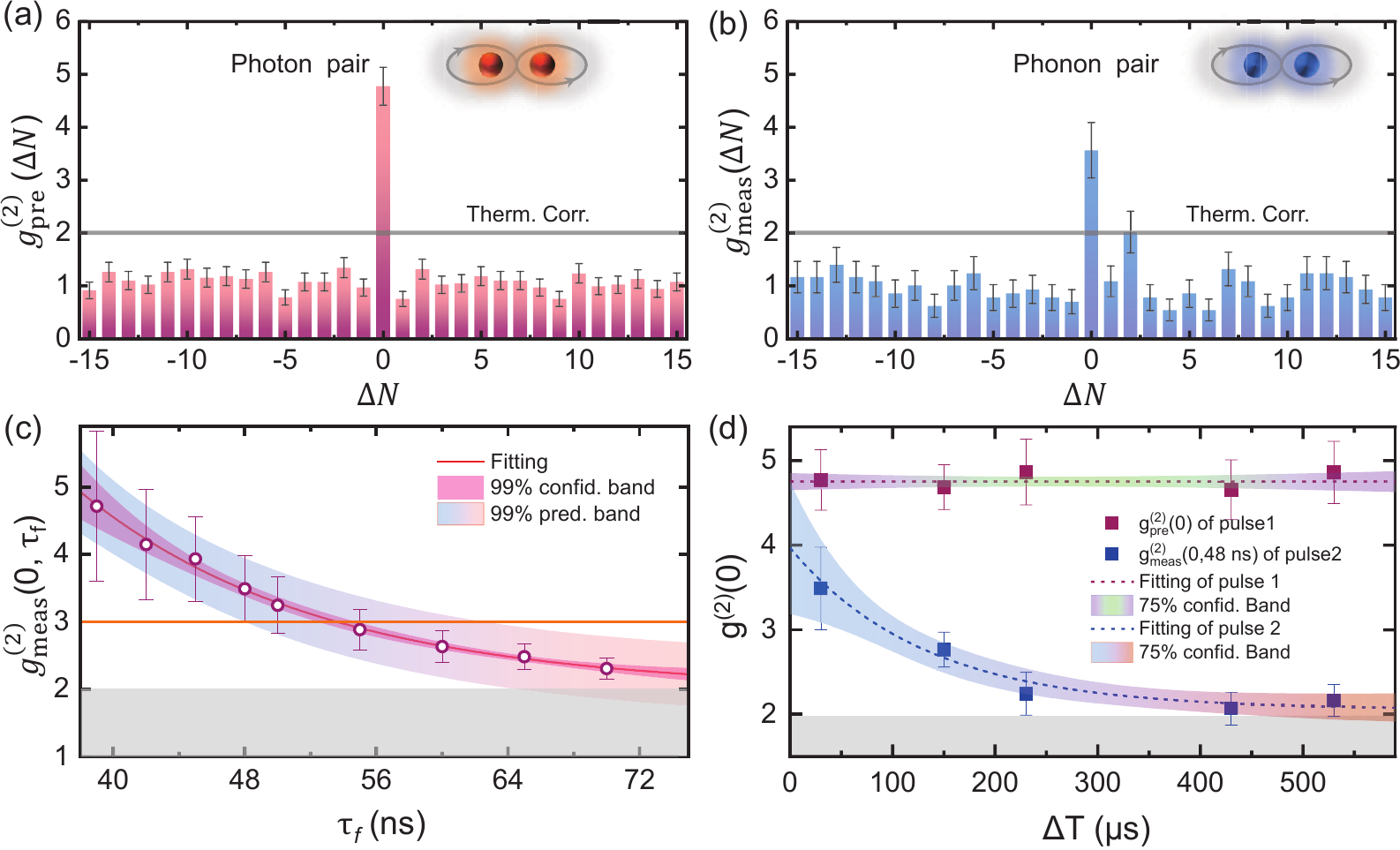}}\caption{\textbf{Experimental demonstration of a phonon pair.} (a-b) Experimental
$g_{\mathrm{pre}}^{\left(2\right)}\left(\Delta N\right)$ and $g_{\mathrm{meas}}^{\left(2\right)}\left(\Delta N\right)$
for the preparation and measurement pulses, respectively. (c) Exponential
decay $g_{\mathrm{meas}}^{\left(2\right)}\left(0,\tau_{f}\right)$
as a function of $\tau_{f}$ with $\Delta T=30\,\mathrm{ns}$. The
red fitting line shows an exponential decay constant of $t_{\mathrm{d1}}\thickapprox15\,\mathrm{ns}$.
The shadow areas indicate the 99\% confidence and prediction intervals.
(d) shows almost identical $g_{\mathrm{pre}}^{\left(2\right)}\left(0\right)$
values while $g_{\mathrm{meas}}^{\left(2\right)}\left(0,48\,\mathrm{ns}\right)$
decays into the thermal state as the time interval $\Delta T$ increases.
The dashed blue line shows the fitted decoherence lifetime $t_{\mathrm{d2}}$
of approximately $132\,\mathrm{ns}$. The shadow area indicates the
75\% confidence interval. All error bars indicate +/\textminus{} one
standard deviation.}

\label{Fig3}
\end{figure*}
A preparation-measurement scheme, repeated every $2\,\mathrm{ms}$,
is implemented, as shown in Fig. \ref{Fig2}(b). First, we prepare
the desired optomechanical quantum state $\left|\Phi\right\rangle _{\mathrm{a,m}}$
by applying the above unitary operator of Eq. \ref{eq:1}. Owing to
the short photon lifetime, the photon pair $\left|20\right\rangle _{\mathrm{a,m}}$
generated with a probability of $1/2$ is rapidly emitted from the
cavity, whereas the phonon pair $\left|02\right\rangle _{\mathrm{a,m}}$
also created with a probability of $1/2$ persists for a moment. The
preparation pulses with a full width at half maximum (FWHM) of $36\,\mathrm{ns}$
create $\left|\Phi\right\rangle _{\mathrm{a,m}}$ with a probability
of $p_{\mathrm{pre}}=p_{\mathrm{b}}p_{\mathrm{r}}\thickapprox2.9\,\%$
(see the Supplementary Information). Consequently, the signal photons
in pulse 1 contain the desired photon pair $\left|20\right\rangle _{\mathrm{a,m}}$
and the inevitable scattered photons from the Stokes and anti-Stokes
processes. The mechanical state is subsequently read out using a $47\,\mathrm{ns}$
(FWHM) red-detuned measurement pulse at various preparation-measurement
time delays, $\Delta T$, which indicate the interval between the
end of the first pulse and the start of the second pulse and satisfy
$1/\gamma_{m}>\Delta T\gg1/\kappa$. The signal photons in pulse 2
contain photons coherently converted from the expected phonon pair
$\left|02\right\rangle _{\mathrm{a,m}}$ and unavoidable anti-Stokes
scattering. The phonon-photon conversion probability $p_{\mathrm{meas}}$
is approximately $24.4\,\%$. The poor preparation probability is
primarily limited by laser heating. The resulting signal photons ($\omega_{\mathrm{a}}$)
are filtered by cascaded filters and launched into two superconducting
nanowire single-photon detectors (SNSPDs) in a Hanbury Brown\textendash Twiss
(HBT) setup, as shown in Fig. \ref{Fig2}(a). Figure \ref{Fig2}(c)
shows the normalized counts for the first and second pulses. These
results are further used for twofold coincidence to verify nonclassical
correlations. The increased counts in pulse 2 at longer $\Delta T$
are attributed to mechanical thermal relaxation (discussed later).

\textbf{\textsl{\emph{Experimental demonstration of a phonon pair.}}}\textbf{\emph{
}}\emph{-}To confirm the generation of nonclassical phonon pairs,
we employ the well-known Cauchy\textendash Schwarz inequality \cite{kuzmich2003generation}.
Two phonons in $\left|02\right\rangle _{\mathrm{a,m}}$ arise from
two correlated pathways, i.e., $\left|02\right\rangle _{\mathrm{a,m}}=\left|0\right\rangle _{\mathrm{a}}\left|1_{\mathrm{b}}1_{\mathrm{r}}\right\rangle _{\mathrm{m}}$:
(I) directly from a single Stokes scattering event, and (II) through
conversion from a previously generated photon via anti-Stokes scattering.
The noncanonical correlations between them are quantified by deriving
a modified Cauchy\textendash Schwarz inequality. For the second-order
correlation function $g^{\left(2\right)}$ of signal photons in this
process, we obtain,
\begin{equation}
g^{\left(2\right)}\leq\frac{\eta^{2}g_{\mathrm{r}}^{\left(2\right)}+g_{\mathrm{b}}^{\left(2\right)}+4\eta\sqrt{g_{\mathrm{r}}^{\left(2\right)}g_{\mathrm{b}}^{\left(2\right)}}}{\left(1+\eta\right)^{2}},\label{eq:3}
\end{equation}
where $g_{\mathrm{r(b)}}^{\left(2\right)}$ is the second-order correlation
of the scattering photons driven solely by the red (blue)-detuned
laser and $\eta$ is related to the ratio of the two Raman scattering
rates. The right-hand side of inequality \ref{eq:3} is maximized
when $\eta=1$ for $g_{\mathrm{r}}^{\left(2\right)}=g_{\mathrm{b}}^{\left(2\right)}$
(see the Supplementary Information). Consequently, for a pair correlation
of classical origin, a stricter boundary applies: $g^{\left(2\right)}\leq\frac{3}{2}\sqrt{g_{\mathrm{r}}^{\left(2\right)}g_{\mathrm{b}}^{\left(2\right)}}$.
Therefore, the nonclassical nature of the phonon pair could be doubtlessly
verified by violation of Eq. \ref{eq:3}.

Figure \ref{Fig3}(a) displays the second-order correlation function
$g_{\mathrm{pre}}^{\left(2\right)}\left(\Delta N\right)$ of signal
photons for the preparation pulse, which is statistically obtained
from numerous repeated pulse sequences of different trials separated
by $\Delta N$ iterations. $g_{\mathrm{pre}}^{\left(2\right)}\left(\Delta N\right)$
quantifies the quantum correlation of the generated photon pair in
pulse 1, while the phonon pair is optionally created at the same rate.
The typical values of $g_{\mathrm{pre}}^{\left(2\right)}\left(\Delta N\neq0\right)$
are close to 1, indicating that signal photons emerging from different
pulse sequences are uncorrelated and that the Cauchy\textendash Schwarz
inequality is fulfilled. In contrast, for pairs emitted from the same
pulse sequence ($\Delta N=0$), $g_{\mathrm{pre}}^{\left(2\right)}\left(0\right)=4.8\pm0.4$,
which significantly violates the classical bound by approximately
$5$ standard deviations, i.e.,
\[
\left[g_{\mathrm{pre}}^{\left(2\right)}\left(0\right)\right]^{2}\nleq9\left[g_{\mathrm{pre,b}}^{\left(2\right)}\left(0\right)g_{\mathrm{pre,r}}^{\left(2\right)}\left(0\right)\right]/4,
\]
where $g_{\mathrm{pre,b}}^{\left(2\right)}\left(0\right)=1.90\pm0.24$,
and $g_{\mathrm{pre,r}}^{\left(2\right)}\left(0\right)=1.88\pm0.42$
(see the Supplementary Information). This violation of the Cauchy\textendash Schwarz
inequality unequivocally demonstrates the nonclassical correlation
of photon pair $\left|20\right\rangle _{\mathrm{a,m}}$.

For state $\Phi_{\mathrm{a,m}}$, phonon pair $\left|02\right\rangle _{\mathrm{a,m}}$
is produced in exactly the same way as optical state $\left|20\right\rangle _{\mathrm{a,m}}$
and possesses the same quantum properties in principle. Direct mapping
of mechanical state $\left|02\right\rangle _{\mathrm{a,m}}$ via a
second pulse, as shown in Fig. \ref{Fig3}(b), would offer more intuitive
evidence of entangled phonon pair generation. However, the $g_{\mathrm{meas}}^{\left(2\right)}\left(0\right)$
for mechanical state $\left|02\right\rangle _{\mathrm{a,m}}$ is worsened
by the initial $n_{\mathrm{th}}$ and additional heating from pulse
2, as shown in Fig. \ref{Fig3}(c). Exponential decay fitting (red
line) reveals a decay constant of $t_{\mathrm{d1}}\thickapprox15\,\mathrm{ns}$,
suggesting accelerated thermal decoherence due to the additional laser
driving.\textcolor{brown}{{} }The shorter the $\tau_{f}$ of pulse 2
is to mitigate laser heating from this pulse, the higher $g_{\mathrm{meas}}^{\left(2\right)}\left(0\right)$
is. Notably, $g_{\mathrm{meas}}^{\left(2\right)}\left(0,39\,\mathrm{ns}\right)=4.7\pm1.1$,
which is beyond the classical bound by approximately $1.5$ standard
deviations, although with larger error bars. \textcolor{black}{Within
the 99\% pre}diction interval, the classical bound is undoubtedly
violated for \textbf{$\tau_{f}$} up to \textbf{$48\,\mathrm{ns}$}.
Therefore, we use click events during the first $48\,\mathrm{ns}$
of pulse 2 to maximize the coincidence counts of the entangled pair.
The resulting\textbf{ $g_{\mathrm{meas}}^{\left(2\right)}\left(\Delta N,48\,\mathrm{ns}\right)$}
is shown in Fig. \ref{Fig3}(b). $g_{\mathrm{meas}}^{\left(2\right)}\left(0,48\,\mathrm{ns}\right)=3.5\pm0.5$
violates the Cauchy\textendash Schwarz inequality,
\[
\left[g_{\mathrm{meas}}^{\left(2\right)}\left(0,48\,\mathrm{ns}\right)\right]^{2}\nleq9\left[g_{\mathrm{meas,r}}^{\left(2\right)}\left(0\right)\right]^{2}/4,
\]
where $g_{\mathrm{meas,r}}^{\left(2\right)}\left(0\right)=1.92\pm0.14$
(see the Supplementary Information), and deviates from the thermal
state by 3.1 standard deviations. Additionally, Fig. \ref{Fig3}(d)
shows that $g_{\mathrm{meas}}^{\left(2\right)}\left(0,48\,\mathrm{ns}\right)$
decreases toward the thermal state as $\Delta T$ increases, whereas
$g_{\mathrm{pre}}^{\left(2\right)}\left(0\right)$ remains almost
unchanged, where the shadow area indicates the 75\% confidence interval.
The fitted decoherence lifetime $t_{\mathrm{d2}}=132\,\mathrm{ns}$
demonstrates that we can store and retrieve the nonclassical mechanical
state $\left|02\right\rangle _{\mathrm{a,m}}$ for an extended time
interval.

\begin{figure}[t]
\centerline{\includegraphics[clip,width=0.9\columnwidth]{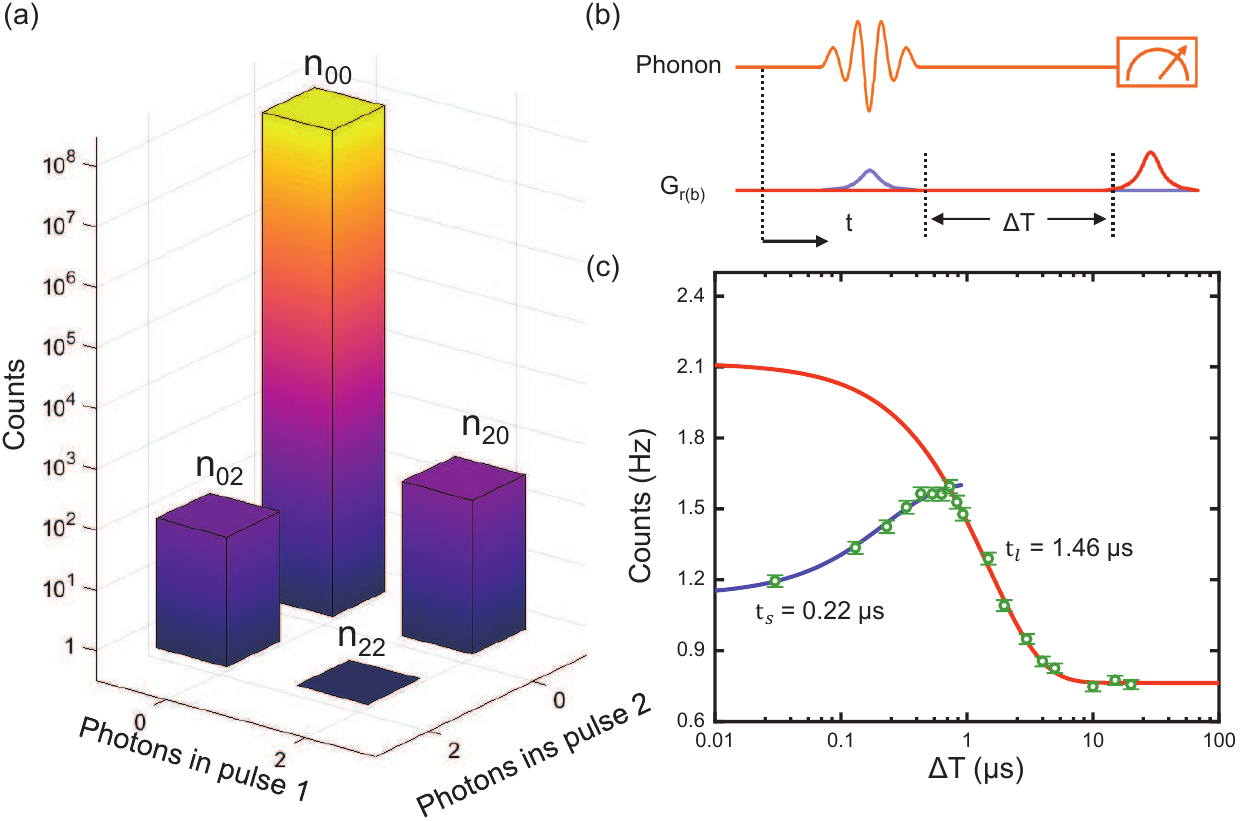}}\caption{(a) Coincidence counts of two-photon click events and zero-photon
click events between pulses 1 and 2, obtained from $N\thickapprox1.4\times10^{8}$
independent equally distributed experiments with $\Delta T=150\,\mathrm{ns}$.
(b) Pump-probe measurement in mechanical breathing mode. A short blue-detuned
pulse (pump) is sent to excite phonon occupation, and then, the mechanical
response is measured via a red-detuned optical probe pulse as a function
of the pump-probe time delay $\Delta T$. (c) The red and blue lines
are the exponential theoretical fits, corresponding to $t_{l}=1.46\,\mathrm{\mu s}$
and $t_{s}=0.22\,\mathrm{\mu s}$, respectively. All error bars indicate
+/\textminus{} one standard deviation.}

\label{Fig4}
\end{figure}

To rule out the possibility of direct excitation of state $\left|22\right\rangle _{\mathrm{a\text{,}m}}$
by the blue-detuned drive, we analyze fourfold coincidence counts
of two-photon click events $n_{22}$ between pulses 1 and 2. Figure
\ref{Fig4}(a) shows the results for $N\thickapprox1.4\times10^{8}$
independent equally distributed experiments with $\Delta T=150\,\mathrm{ns}$.
We observe $n_{22}=0$, whereas the coincidence counts between two-photon
click events in pulse 1 (2) and zero-photon click events in pulse
2 (1) are significantly greater. This result indicates negligible
excitation of state $\left|22\right\rangle _{\mathrm{a\text{,}m}}$.
Moreover, the pair events generated in the two pulses exhibit mutual
repulsion, which is consistent with the behavior of the declared state
$\left|\Phi\right\rangle _{\mathrm{a,m}}=\left(\left|20\right\rangle _{\mathrm{a,m}}+\left|02\right\rangle _{\mathrm{a,m}}\right)/\sqrt{2}$.
Consequently, the use of pair events $\left|20\right\rangle _{\mathrm{a,m}}$
in pulse 1 allows post-selection of a mechanical vacuum state. Conversely,
the detection of zero-photon events $\left|00\right\rangle _{\mathrm{a,m}}$
in pulse 1 projects state $\left|\varphi\right\rangle _{\mathrm{a,m}}$
to
\begin{equation}
\left|\varphi'\right\rangle _{\mathrm{a,m}}=\left|00\right\rangle _{\mathrm{a,m}}+\sqrt{p_{\mathrm{b}}p_{\mathrm{r}}}\left|02\right\rangle _{\mathrm{a,m}}+O(p_{\mathrm{b}}),
\end{equation}
effectively doubling the phonon pair probability. \textcolor{black}{In
this scenario, the generation rate of the phonon pair is $\varpropto$
$p_{\mathrm{b}}$, surpassing the $p_{\mathrm{b}}^{2}$ rate achievable
with only ``two-mode squeezed'' interaction.}

To identify the limitations on the storage time of the phonon pair,
we measure the mechanical mode decay lifetime using a pump-probe scheme,
as shown in Fig. \ref{Fig4}(b). A short blue-detuned pulse (pump)
excites phonon occupation, and then the mechanical response is measured
by a red-detuned probe pulse as a function of the pump\textendash probe
time delay ($\Delta T$). Figure \ref{Fig4}(c) displays the results.
The long-term mechanical response ($\Delta T\geqslant1\,\mathrm{\mu s}$)
well fits a simple exponential decay (red line) with a decay time
constant ($t_{l}$) of $1.46\,\mathrm{\mu s}$, which is consistent
with the mechanical $Q_{\mathrm{m}}$ of $4.8\times10^{4}$. However,
the short-term mechanical response ($\Delta T<1\,\mathrm{\mu s}$)
shows mechanical thermal relaxation due to laser heating. The mechanical
thermal relaxation time can be fitted to another exponential curve
with a time constant of $t_{s}=0.22\,\mathrm{\mu s}$ (blue line).
The scale of $t_{s}$ is consistent with $t_{\mathrm{d2}}$ and with
the increasing count rates in Fig. \ref{Fig2}(c). Therefore, the
quantum properties of our phonon pair are limited primarily by laser-induced
mechanical thermal relaxation, which drives the mechanical system
toward the thermal state.

Unlike previous work on quantum squeezing \cite{wollman2015quantum}
and quantum nondemolition measurement \cite{lei2016quantum} of continuous
variables (typically employing homodyne or heterodyne detection of
two tones of a continuous pump), our experiment utilizes a pulse scheme.
This approach enables access to the evolution of discrete quantum
variables using the equivalent high-order optomechanical nonlinearity,
offering the potential to control the evolutionary phases of individual
states through the driving intensity and duration. Here, we have generated
quantum-correlated phonon pairs via optical control based on the quadratic
term in the time-evolution operator $U\left(t\right)$. The dominant
limitation originates from the heating induced by both the preparation
and measurement pulses, and even better results can be expected by
further reducing the laser heating, such as when 2D OMC structures
are used\cite{chen2024bandwidth}.

\textbf{\emph{Conclusion.}}\emph{ -} In summary, this work demonstrates,
for the first time, the generation and verification of quantum correlated
phonon pairs via optical control based on effective optomechanical
SFWM and violation of the Cauchy\textendash Schwarz inequality. A
pulse scheme enables access to equivalent high-order optomechanical
nonlinearity\textendash \textendash primarily an effective optomechanical
SFWM process here\textendash \textendash facilitating the creation
of an optomechanical NOON state and the conditional generation of
a quantum-correlated phonon pair. Our experiments demonstrate violations
of the Cauchy\textendash Schwarz inequality by over 5 standard deviations
for the photon pair and 1.5 standard deviations for the phonon pair.
The decoherence lifetime of the phonon pair is 132 ns. This research
opens new avenues for experimentally exploring effective high-order
nonlinearity in quantum optomechanics, and expands the available techniques
for quantum optomechanical manipulation.

\bibliographystyle{Microcavity}

\begin{thebibliography}{39}%
\makeatletter
\providecommand \@ifxundefined [1]{%
 \@ifx{#1\undefined}
}%
\providecommand \@ifnum [1]{%
 \ifnum #1\expandafter \@firstoftwo
 \else \expandafter \@secondoftwo
 \fi
}%
\providecommand \@ifx [1]{%
 \ifx #1\expandafter \@firstoftwo
 \else \expandafter \@secondoftwo
 \fi
}%
\providecommand \natexlab [1]{#1}%
\providecommand \enquote  [1]{``#1''}%
\providecommand \bibnamefont  [1]{#1}%
\providecommand \bibfnamefont [1]{#1}%
\providecommand \citenamefont [1]{#1}%
\providecommand \href@noop [0]{\@secondoftwo}%
\providecommand \href [0]{\begingroup \@sanitize@url \@href}%
\providecommand \@href[1]{\@@startlink{#1}\@@href}%
\providecommand \@@href[1]{\endgroup#1\@@endlink}%
\providecommand \@sanitize@url [0]{\catcode `\\12\catcode `\$12\catcode
  `\&12\catcode `\#12\catcode `\^12\catcode `\_12\catcode `\%12\relax}%
\providecommand \@@startlink[1]{}%
\providecommand \@@endlink[0]{}%
\providecommand \url  [0]{\begingroup\@sanitize@url \@url }%
\providecommand \@url [1]{\endgroup\@href {#1}{\urlprefix }}%
\providecommand \urlprefix  [0]{URL }%
\providecommand \Eprint [0]{\href }%
\providecommand \doibase [0]{http://dx.doi.org/}%
\providecommand \selectlanguage [0]{\@gobble}%
\providecommand \bibinfo  [0]{\@secondoftwo}%
\providecommand \bibfield  [0]{\@secondoftwo}%
\providecommand \translation [1]{[#1]}%
\providecommand \BibitemOpen [0]{}%
\providecommand \bibitemStop [0]{}%
\providecommand \bibitemNoStop [0]{.\EOS\space}%
\providecommand \EOS [0]{\spacefactor3000\relax}%
\providecommand \BibitemShut  [1]{\csname bibitem#1\endcsname}%
\let\auto@bib@innerbib\@empty
\bibitem [{\citenamefont {Kimble}(2008)}]{kimble2008quantum}%
  \BibitemOpen
  \bibfield  {author} {\bibinfo {author} {\bibfnamefont {H.~J.}\ \bibnamefont
  {Kimble}},\ }\bibfield  {title} {\enquote {\bibinfo {title} {The quantum
  internet},}\ }\href@noop {} {\bibfield  {journal} {\bibinfo  {journal}
  {Nature}\ }\textbf {\bibinfo {volume} {453}},\ \bibinfo {pages} {1023}
  (\bibinfo {year} {2008})}\BibitemShut {NoStop}%
\bibitem [{\citenamefont {Walther}\ \emph {et~al.}(2005)\citenamefont
  {Walther}, \citenamefont {Resch}, \citenamefont {Rudolph}, \citenamefont
  {Schenck}, \citenamefont {Weinfurter}, \citenamefont {Vedral}, \citenamefont
  {Aspelmeyer},\ and\ \citenamefont {Zeilinger}}]{walther2005experimental}%
  \BibitemOpen
  \bibfield  {author} {\bibinfo {author} {\bibfnamefont {P.}~\bibnamefont
  {Walther}}, \bibinfo {author} {\bibfnamefont {K.~J.}\ \bibnamefont {Resch}},
  \bibinfo {author} {\bibfnamefont {T.}~\bibnamefont {Rudolph}}, \bibinfo
  {author} {\bibfnamefont {E.}~\bibnamefont {Schenck}}, \bibinfo {author}
  {\bibfnamefont {H.}~\bibnamefont {Weinfurter}}, \bibinfo {author}
  {\bibfnamefont {V.}~\bibnamefont {Vedral}}, \bibinfo {author} {\bibfnamefont
  {M.}~\bibnamefont {Aspelmeyer}}, \ and\ \bibinfo {author} {\bibfnamefont
  {A.}~\bibnamefont {Zeilinger}},\ }\bibfield  {title} {\enquote {\bibinfo
  {title} {Experimental one-way quantum computing},}\ }\href@noop {} {\bibfield
   {journal} {\bibinfo  {journal} {Nature}\ }\textbf {\bibinfo {volume}
  {434}},\ \bibinfo {pages} {169} (\bibinfo {year} {2005})}\BibitemShut
  {NoStop}%
\bibitem [{\citenamefont {Giovannetti}\ \emph {et~al.}(2004)\citenamefont
  {Giovannetti}, \citenamefont {Lloyd},\ and\ \citenamefont
  {Maccone}}]{giovannetti2004quantum}%
  \BibitemOpen
  \bibfield  {author} {\bibinfo {author} {\bibfnamefont {V.}~\bibnamefont
  {Giovannetti}}, \bibinfo {author} {\bibfnamefont {S.}~\bibnamefont {Lloyd}},
  \ and\ \bibinfo {author} {\bibfnamefont {L.}~\bibnamefont {Maccone}},\
  }\bibfield  {title} {\enquote {\bibinfo {title} {Quantum-enhanced
  measurements: beating the standard quantum limit},}\ }\href@noop {}
  {\bibfield  {journal} {\bibinfo  {journal} {Science}\ }\textbf {\bibinfo
  {volume} {306}},\ \bibinfo {pages} {1330} (\bibinfo {year}
  {2004})}\BibitemShut {NoStop}%
\bibitem [{\citenamefont {O'brien}\ \emph {et~al.}(2009)\citenamefont
  {O'brien}, \citenamefont {Furusawa},\ and\ \citenamefont
  {Vu{\v{c}}kovi{\'c}}}]{o2009photonic}%
  \BibitemOpen
  \bibfield  {author} {\bibinfo {author} {\bibfnamefont {J.~L.}\ \bibnamefont
  {O'brien}}, \bibinfo {author} {\bibfnamefont {A.}~\bibnamefont {Furusawa}}, \
  and\ \bibinfo {author} {\bibfnamefont {J.}~\bibnamefont
  {Vu{\v{c}}kovi{\'c}}},\ }\bibfield  {title} {\enquote {\bibinfo {title}
  {Photonic quantum technologies},}\ }\href@noop {} {\bibfield  {journal}
  {\bibinfo  {journal} {Nature Photonics}\ }\textbf {\bibinfo {volume} {3}},\
  \bibinfo {pages} {687} (\bibinfo {year} {2009})}\BibitemShut {NoStop}%
\bibitem [{\citenamefont {Pezze}\ \emph {et~al.}(2018)\citenamefont {Pezze},
  \citenamefont {Smerzi}, \citenamefont {Oberthaler}, \citenamefont {Schmied},\
  and\ \citenamefont {Treutlein}}]{pezze2018quantum}%
  \BibitemOpen
  \bibfield  {author} {\bibinfo {author} {\bibfnamefont {L.}~\bibnamefont
  {Pezze}}, \bibinfo {author} {\bibfnamefont {A.}~\bibnamefont {Smerzi}},
  \bibinfo {author} {\bibfnamefont {M.~K.}\ \bibnamefont {Oberthaler}},
  \bibinfo {author} {\bibfnamefont {R.}~\bibnamefont {Schmied}}, \ and\
  \bibinfo {author} {\bibfnamefont {P.}~\bibnamefont {Treutlein}},\ }\bibfield
  {title} {\enquote {\bibinfo {title} {Quantum metrology with nonclassical
  states of atomic ensembles},}\ }\href@noop {} {\bibfield  {journal} {\bibinfo
   {journal} {Reviews of Modern Physics}\ }\textbf {\bibinfo {volume} {90}},\
  \bibinfo {pages} {035005} (\bibinfo {year} {2018})}\BibitemShut {NoStop}%
\bibitem [{\citenamefont {Giustina}\ \emph {et~al.}(2015)\citenamefont
  {Giustina}, \citenamefont {Versteegh}, \citenamefont {Wengerowsky},
  \citenamefont {Handsteiner}, \citenamefont {Hochrainer}, \citenamefont
  {Phelan}, \citenamefont {Steinlechner}, \citenamefont {Kofler}, \citenamefont
  {Larsson}, \citenamefont {Abell{\'a}n} \emph
  {et~al.}}]{giustina2015significant}%
  \BibitemOpen
  \bibfield  {author} {\bibinfo {author} {\bibfnamefont {M.}~\bibnamefont
  {Giustina}}, \bibinfo {author} {\bibfnamefont {M.~A.}\ \bibnamefont
  {Versteegh}}, \bibinfo {author} {\bibfnamefont {S.}~\bibnamefont
  {Wengerowsky}}, \bibinfo {author} {\bibfnamefont {J.}~\bibnamefont
  {Handsteiner}}, \bibinfo {author} {\bibfnamefont {A.}~\bibnamefont
  {Hochrainer}}, \bibinfo {author} {\bibfnamefont {K.}~\bibnamefont {Phelan}},
  \bibinfo {author} {\bibfnamefont {F.}~\bibnamefont {Steinlechner}}, \bibinfo
  {author} {\bibfnamefont {J.}~\bibnamefont {Kofler}}, \bibinfo {author}
  {\bibfnamefont {J.-{\AA}.}\ \bibnamefont {Larsson}}, \bibinfo {author}
  {\bibfnamefont {C.}~\bibnamefont {Abell{\'a}n}},  \emph {et~al.},\ }\bibfield
   {title} {\enquote {\bibinfo {title} {Significant-loophole-free test of
  bell¡¯s theorem with entangled photons},}\ }\href@noop {} {\bibfield
  {journal} {\bibinfo  {journal} {Physical review letters}\ }\textbf {\bibinfo
  {volume} {115}},\ \bibinfo {pages} {250401} (\bibinfo {year}
  {2015})}\BibitemShut {NoStop}%
\bibitem [{\citenamefont {Sangouard}\ \emph {et~al.}(2011)\citenamefont
  {Sangouard}, \citenamefont {Simon}, \citenamefont {De~Riedmatten},\ and\
  \citenamefont {Gisin}}]{sangouard2011quantum}%
  \BibitemOpen
  \bibfield  {author} {\bibinfo {author} {\bibfnamefont {N.}~\bibnamefont
  {Sangouard}}, \bibinfo {author} {\bibfnamefont {C.}~\bibnamefont {Simon}},
  \bibinfo {author} {\bibfnamefont {H.}~\bibnamefont {De~Riedmatten}}, \ and\
  \bibinfo {author} {\bibfnamefont {N.}~\bibnamefont {Gisin}},\ }\bibfield
  {title} {\enquote {\bibinfo {title} {Quantum repeaters based on atomic
  ensembles and linear optics},}\ }\href@noop {} {\bibfield  {journal}
  {\bibinfo  {journal} {Reviews of Modern Physics}\ }\textbf {\bibinfo {volume}
  {83}},\ \bibinfo {pages} {33} (\bibinfo {year} {2011})}\BibitemShut {NoStop}%
\bibitem [{\citenamefont {Liu}\ \emph {et~al.}(2021)\citenamefont {Liu},
  \citenamefont {Hu}, \citenamefont {Li}, \citenamefont {Li}, \citenamefont
  {Li}, \citenamefont {Liang}, \citenamefont {Zhou}, \citenamefont {Li},\ and\
  \citenamefont {Guo}}]{liu2021heralded}%
  \BibitemOpen
  \bibfield  {author} {\bibinfo {author} {\bibfnamefont {X.}~\bibnamefont
  {Liu}}, \bibinfo {author} {\bibfnamefont {J.}~\bibnamefont {Hu}}, \bibinfo
  {author} {\bibfnamefont {Z.-F.}\ \bibnamefont {Li}}, \bibinfo {author}
  {\bibfnamefont {X.}~\bibnamefont {Li}}, \bibinfo {author} {\bibfnamefont
  {P.-Y.}\ \bibnamefont {Li}}, \bibinfo {author} {\bibfnamefont {P.-J.}\
  \bibnamefont {Liang}}, \bibinfo {author} {\bibfnamefont {Z.-Q.}\ \bibnamefont
  {Zhou}}, \bibinfo {author} {\bibfnamefont {C.-F.}\ \bibnamefont {Li}}, \ and\
  \bibinfo {author} {\bibfnamefont {G.-C.}\ \bibnamefont {Guo}},\ }\bibfield
  {title} {\enquote {\bibinfo {title} {Heralded entanglement distribution
  between two absorptive quantum memories},}\ }\href@noop {} {\bibfield
  {journal} {\bibinfo  {journal} {Nature}\ }\textbf {\bibinfo {volume} {594}},\
  \bibinfo {pages} {41} (\bibinfo {year} {2021})}\BibitemShut {NoStop}%
\bibitem [{\citenamefont {Wallucks}\ \emph {et~al.}(2020)\citenamefont
  {Wallucks}, \citenamefont {Marinkovi{\'c}}, \citenamefont {Hensen},
  \citenamefont {Stockill},\ and\ \citenamefont
  {Gr{\"o}blacher}}]{wallucks2020quantum}%
  \BibitemOpen
  \bibfield  {author} {\bibinfo {author} {\bibfnamefont {A.}~\bibnamefont
  {Wallucks}}, \bibinfo {author} {\bibfnamefont {I.}~\bibnamefont
  {Marinkovi{\'c}}}, \bibinfo {author} {\bibfnamefont {B.}~\bibnamefont
  {Hensen}}, \bibinfo {author} {\bibfnamefont {R.}~\bibnamefont {Stockill}}, \
  and\ \bibinfo {author} {\bibfnamefont {S.}~\bibnamefont {Gr{\"o}blacher}},\
  }\bibfield  {title} {\enquote {\bibinfo {title} {A quantum memory at telecom
  wavelengths},}\ }\href@noop {} {\bibfield  {journal} {\bibinfo  {journal}
  {Nature Physics}\ }\textbf {\bibinfo {volume} {16}},\ \bibinfo {pages} {772}
  (\bibinfo {year} {2020})}\BibitemShut {NoStop}%
\bibitem [{\citenamefont {Ac{\'\i}n}\ \emph {et~al.}(2007)\citenamefont
  {Ac{\'\i}n}, \citenamefont {Brunner}, \citenamefont {Gisin}, \citenamefont
  {Massar}, \citenamefont {Pironio},\ and\ \citenamefont
  {Scarani}}]{acin2007device}%
  \BibitemOpen
  \bibfield  {author} {\bibinfo {author} {\bibfnamefont {A.}~\bibnamefont
  {Ac{\'\i}n}}, \bibinfo {author} {\bibfnamefont {N.}~\bibnamefont {Brunner}},
  \bibinfo {author} {\bibfnamefont {N.}~\bibnamefont {Gisin}}, \bibinfo
  {author} {\bibfnamefont {S.}~\bibnamefont {Massar}}, \bibinfo {author}
  {\bibfnamefont {S.}~\bibnamefont {Pironio}}, \ and\ \bibinfo {author}
  {\bibfnamefont {V.}~\bibnamefont {Scarani}},\ }\bibfield  {title} {\enquote
  {\bibinfo {title} {Device-independent security of quantum cryptography
  against collective attacks},}\ }\href@noop {} {\bibfield  {journal} {\bibinfo
   {journal} {Physical Review Letters}\ }\textbf {\bibinfo {volume} {98}},\
  \bibinfo {pages} {230501} (\bibinfo {year} {2007})}\BibitemShut {NoStop}%
\bibitem [{\citenamefont {Pan}\ \emph {et~al.}(2012)\citenamefont {Pan},
  \citenamefont {Chen}, \citenamefont {Lu}, \citenamefont {Weinfurter},
  \citenamefont {Zeilinger},\ and\ \citenamefont
  {{\.Z}ukowski}}]{pan2012multiphoton}%
  \BibitemOpen
  \bibfield  {author} {\bibinfo {author} {\bibfnamefont {J.-W.}\ \bibnamefont
  {Pan}}, \bibinfo {author} {\bibfnamefont {Z.-B.}\ \bibnamefont {Chen}},
  \bibinfo {author} {\bibfnamefont {C.-Y.}\ \bibnamefont {Lu}}, \bibinfo
  {author} {\bibfnamefont {H.}~\bibnamefont {Weinfurter}}, \bibinfo {author}
  {\bibfnamefont {A.}~\bibnamefont {Zeilinger}}, \ and\ \bibinfo {author}
  {\bibfnamefont {M.}~\bibnamefont {{\.Z}ukowski}},\ }\bibfield  {title}
  {\enquote {\bibinfo {title} {Multiphoton entanglement and interferometry},}\
  }\href@noop {} {\bibfield  {journal} {\bibinfo  {journal} {Reviews of Modern
  Physics}\ }\textbf {\bibinfo {volume} {84}},\ \bibinfo {pages} {777}
  (\bibinfo {year} {2012})}\BibitemShut {NoStop}%
\bibitem [{\citenamefont {Sheremet}\ \emph {et~al.}(2023)\citenamefont
  {Sheremet}, \citenamefont {Petrov}, \citenamefont {Iorsh}, \citenamefont
  {Poshakinskiy},\ and\ \citenamefont {Poddubny}}]{sheremet2023waveguide}%
  \BibitemOpen
  \bibfield  {author} {\bibinfo {author} {\bibfnamefont {A.~S.}\ \bibnamefont
  {Sheremet}}, \bibinfo {author} {\bibfnamefont {M.~I.}\ \bibnamefont
  {Petrov}}, \bibinfo {author} {\bibfnamefont {I.~V.}\ \bibnamefont {Iorsh}},
  \bibinfo {author} {\bibfnamefont {A.~V.}\ \bibnamefont {Poshakinskiy}}, \
  and\ \bibinfo {author} {\bibfnamefont {A.~N.}\ \bibnamefont {Poddubny}},\
  }\bibfield  {title} {\enquote {\bibinfo {title} {Waveguide quantum
  electrodynamics: Collective radiance and photon-photon correlations},}\
  }\href@noop {} {\bibfield  {journal} {\bibinfo  {journal} {Reviews of Modern
  Physics}\ }\textbf {\bibinfo {volume} {95}},\ \bibinfo {pages} {015002}
  (\bibinfo {year} {2023})}\BibitemShut {NoStop}%
\bibitem [{\citenamefont {Meesala}\ \emph {et~al.}(2024)\citenamefont
  {Meesala}, \citenamefont {Wood}, \citenamefont {Lake}, \citenamefont
  {Chiappina}, \citenamefont {Zhong}, \citenamefont {Beyer}, \citenamefont
  {Shaw}, \citenamefont {Jiang},\ and\ \citenamefont
  {Painter}}]{meesala2023non}%
  \BibitemOpen
  \bibfield  {author} {\bibinfo {author} {\bibfnamefont {S.}~\bibnamefont
  {Meesala}}, \bibinfo {author} {\bibfnamefont {S.}~\bibnamefont {Wood}},
  \bibinfo {author} {\bibfnamefont {D.}~\bibnamefont {Lake}}, \bibinfo {author}
  {\bibfnamefont {P.}~\bibnamefont {Chiappina}}, \bibinfo {author}
  {\bibfnamefont {C.}~\bibnamefont {Zhong}}, \bibinfo {author} {\bibfnamefont
  {A.~D.}\ \bibnamefont {Beyer}}, \bibinfo {author} {\bibfnamefont {M.~D.}\
  \bibnamefont {Shaw}}, \bibinfo {author} {\bibfnamefont {L.}~\bibnamefont
  {Jiang}}, \ and\ \bibinfo {author} {\bibfnamefont {O.}~\bibnamefont
  {Painter}},\ }\bibfield  {title} {\enquote {\bibinfo {title} {Non-classical
  microwave--optical photon pair generation with a chip-scale transducer},}\
  }\href@noop {} {\bibfield  {journal} {\bibinfo  {journal} {Nature Physics}\
  ,\ \bibinfo {pages} {1}} (\bibinfo {year} {2024})}\BibitemShut {NoStop}%
\bibitem [{\citenamefont {Volz}\ \emph {et~al.}(2006)\citenamefont {Volz},
  \citenamefont {Weber}, \citenamefont {Schlenk}, \citenamefont {Rosenfeld},
  \citenamefont {Vrana}, \citenamefont {Saucke}, \citenamefont {Kurtsiefer},\
  and\ \citenamefont {Weinfurter}}]{volz2006observation}%
  \BibitemOpen
  \bibfield  {author} {\bibinfo {author} {\bibfnamefont {J.}~\bibnamefont
  {Volz}}, \bibinfo {author} {\bibfnamefont {M.}~\bibnamefont {Weber}},
  \bibinfo {author} {\bibfnamefont {D.}~\bibnamefont {Schlenk}}, \bibinfo
  {author} {\bibfnamefont {W.}~\bibnamefont {Rosenfeld}}, \bibinfo {author}
  {\bibfnamefont {J.}~\bibnamefont {Vrana}}, \bibinfo {author} {\bibfnamefont
  {K.}~\bibnamefont {Saucke}}, \bibinfo {author} {\bibfnamefont
  {C.}~\bibnamefont {Kurtsiefer}}, \ and\ \bibinfo {author} {\bibfnamefont
  {H.}~\bibnamefont {Weinfurter}},\ }\bibfield  {title} {\enquote {\bibinfo
  {title} {Observation of entanglement of a single photon with a trapped
  atom},}\ }\href@noop {} {\bibfield  {journal} {\bibinfo  {journal} {Physical
  review letters}\ }\textbf {\bibinfo {volume} {96}},\ \bibinfo {pages}
  {030404} (\bibinfo {year} {2006})}\BibitemShut {NoStop}%
\bibitem [{\citenamefont {Gao}\ \emph {et~al.}(2012)\citenamefont {Gao},
  \citenamefont {Fallahi}, \citenamefont {Togan}, \citenamefont
  {Miguel-S{\'a}nchez},\ and\ \citenamefont {Imamoglu}}]{gao2012observation}%
  \BibitemOpen
  \bibfield  {author} {\bibinfo {author} {\bibfnamefont {W.}~\bibnamefont
  {Gao}}, \bibinfo {author} {\bibfnamefont {P.}~\bibnamefont {Fallahi}},
  \bibinfo {author} {\bibfnamefont {E.}~\bibnamefont {Togan}}, \bibinfo
  {author} {\bibfnamefont {J.}~\bibnamefont {Miguel-S{\'a}nchez}}, \ and\
  \bibinfo {author} {\bibfnamefont {A.}~\bibnamefont {Imamoglu}},\ }\bibfield
  {title} {\enquote {\bibinfo {title} {Observation of entanglement between a
  quantum dot spin and a single photon},}\ }\href@noop {} {\bibfield  {journal}
  {\bibinfo  {journal} {Nature}\ }\textbf {\bibinfo {volume} {491}},\ \bibinfo
  {pages} {426} (\bibinfo {year} {2012})}\BibitemShut {NoStop}%
\bibitem [{\citenamefont {Riedinger}\ \emph {et~al.}(2016)\citenamefont
  {Riedinger}, \citenamefont {Hong}, \citenamefont {Norte}, \citenamefont
  {Slater}, \citenamefont {Shang}, \citenamefont {Krause}, \citenamefont
  {Anant}, \citenamefont {Aspelmeyer},\ and\ \citenamefont
  {Gr{\"o}blacher}}]{riedinger2016non}%
  \BibitemOpen
  \bibfield  {author} {\bibinfo {author} {\bibfnamefont {R.}~\bibnamefont
  {Riedinger}}, \bibinfo {author} {\bibfnamefont {S.}~\bibnamefont {Hong}},
  \bibinfo {author} {\bibfnamefont {R.~A.}\ \bibnamefont {Norte}}, \bibinfo
  {author} {\bibfnamefont {J.~A.}\ \bibnamefont {Slater}}, \bibinfo {author}
  {\bibfnamefont {J.}~\bibnamefont {Shang}}, \bibinfo {author} {\bibfnamefont
  {A.~G.}\ \bibnamefont {Krause}}, \bibinfo {author} {\bibfnamefont
  {V.}~\bibnamefont {Anant}}, \bibinfo {author} {\bibfnamefont
  {M.}~\bibnamefont {Aspelmeyer}}, \ and\ \bibinfo {author} {\bibfnamefont
  {S.}~\bibnamefont {Gr{\"o}blacher}},\ }\bibfield  {title} {\enquote {\bibinfo
  {title} {Non-classical correlations between single photons and phonons from a
  mechanical oscillator},}\ }\href@noop {} {\bibfield  {journal} {\bibinfo
  {journal} {Nature}\ }\textbf {\bibinfo {volume} {530}},\ \bibinfo {pages}
  {313} (\bibinfo {year} {2016})}\BibitemShut {NoStop}%
\bibitem [{\citenamefont {Feist}\ \emph {et~al.}(2022)\citenamefont {Feist},
  \citenamefont {Huang}, \citenamefont {Arend}, \citenamefont {Yang},
  \citenamefont {Henke}, \citenamefont {Raja}, \citenamefont {Kappert},
  \citenamefont {Wang}, \citenamefont {Louren{\c{c}}o-Martins}, \citenamefont
  {Qiu} \emph {et~al.}}]{feist2022cavity}%
  \BibitemOpen
  \bibfield  {author} {\bibinfo {author} {\bibfnamefont {A.}~\bibnamefont
  {Feist}}, \bibinfo {author} {\bibfnamefont {G.}~\bibnamefont {Huang}},
  \bibinfo {author} {\bibfnamefont {G.}~\bibnamefont {Arend}}, \bibinfo
  {author} {\bibfnamefont {Y.}~\bibnamefont {Yang}}, \bibinfo {author}
  {\bibfnamefont {J.-W.}\ \bibnamefont {Henke}}, \bibinfo {author}
  {\bibfnamefont {A.~S.}\ \bibnamefont {Raja}}, \bibinfo {author}
  {\bibfnamefont {F.~J.}\ \bibnamefont {Kappert}}, \bibinfo {author}
  {\bibfnamefont {R.~N.}\ \bibnamefont {Wang}}, \bibinfo {author}
  {\bibfnamefont {H.}~\bibnamefont {Louren{\c{c}}o-Martins}}, \bibinfo {author}
  {\bibfnamefont {Z.}~\bibnamefont {Qiu}},  \emph {et~al.},\ }\bibfield
  {title} {\enquote {\bibinfo {title} {Cavity-mediated electron-photon
  pairs},}\ }\href@noop {} {\bibfield  {journal} {\bibinfo  {journal}
  {Science}\ }\textbf {\bibinfo {volume} {377}},\ \bibinfo {pages} {777}
  (\bibinfo {year} {2022})}\BibitemShut {NoStop}%
\bibitem [{\citenamefont {Clerk}\ \emph {et~al.}(2020)\citenamefont {Clerk},
  \citenamefont {Lehnert}, \citenamefont {Bertet}, \citenamefont {Petta},\ and\
  \citenamefont {Nakamura}}]{clerk2020hybrid}%
  \BibitemOpen
  \bibfield  {author} {\bibinfo {author} {\bibfnamefont {A.}~\bibnamefont
  {Clerk}}, \bibinfo {author} {\bibfnamefont {K.}~\bibnamefont {Lehnert}},
  \bibinfo {author} {\bibfnamefont {P.}~\bibnamefont {Bertet}}, \bibinfo
  {author} {\bibfnamefont {J.}~\bibnamefont {Petta}}, \ and\ \bibinfo {author}
  {\bibfnamefont {Y.}~\bibnamefont {Nakamura}},\ }\bibfield  {title} {\enquote
  {\bibinfo {title} {Hybrid quantum systems with circuit quantum
  electrodynamics},}\ }\href@noop {} {\bibfield  {journal} {\bibinfo  {journal}
  {Nature Physics}\ }\textbf {\bibinfo {volume} {16}},\ \bibinfo {pages} {257}
  (\bibinfo {year} {2020})}\BibitemShut {NoStop}%
\bibitem [{\citenamefont {Shen}\ \emph {et~al.}(2022)\citenamefont {Shen},
  \citenamefont {Xu}, \citenamefont {Zhang}, \citenamefont {Zhang},
  \citenamefont {Wang}, \citenamefont {Chai}, \citenamefont {Zou},
  \citenamefont {Guo},\ and\ \citenamefont {Dong}}]{shen2022coherent}%
  \BibitemOpen
  \bibfield  {author} {\bibinfo {author} {\bibfnamefont {Z.}~\bibnamefont
  {Shen}}, \bibinfo {author} {\bibfnamefont {G.-T.}\ \bibnamefont {Xu}},
  \bibinfo {author} {\bibfnamefont {M.}~\bibnamefont {Zhang}}, \bibinfo
  {author} {\bibfnamefont {Y.-L.}\ \bibnamefont {Zhang}}, \bibinfo {author}
  {\bibfnamefont {Y.}~\bibnamefont {Wang}}, \bibinfo {author} {\bibfnamefont
  {C.-Z.}\ \bibnamefont {Chai}}, \bibinfo {author} {\bibfnamefont {C.-L.}\
  \bibnamefont {Zou}}, \bibinfo {author} {\bibfnamefont {G.-C.}\ \bibnamefont
  {Guo}}, \ and\ \bibinfo {author} {\bibfnamefont {C.-H.}\ \bibnamefont
  {Dong}},\ }\bibfield  {title} {\enquote {\bibinfo {title} {Coherent coupling
  between phonons, magnons, and photons},}\ }\href@noop {} {\bibfield
  {journal} {\bibinfo  {journal} {Physical Review Letters}\ }\textbf {\bibinfo
  {volume} {129}},\ \bibinfo {pages} {243601} (\bibinfo {year}
  {2022})}\BibitemShut {NoStop}%
\bibitem [{\citenamefont {Aspelmeyer}\ \emph {et~al.}(2014)\citenamefont
  {Aspelmeyer}, \citenamefont {Kippenberg},\ and\ \citenamefont
  {Marquardt}}]{aspelmeyer2014cavity}%
  \BibitemOpen
  \bibfield  {author} {\bibinfo {author} {\bibfnamefont {M.}~\bibnamefont
  {Aspelmeyer}}, \bibinfo {author} {\bibfnamefont {T.~J.}\ \bibnamefont
  {Kippenberg}}, \ and\ \bibinfo {author} {\bibfnamefont {F.}~\bibnamefont
  {Marquardt}},\ }\bibfield  {title} {\enquote {\bibinfo {title} {Cavity
  optomechanics},}\ }\href@noop {} {\bibfield  {journal} {\bibinfo  {journal}
  {Reviews of Modern Physics}\ }\textbf {\bibinfo {volume} {86}},\ \bibinfo
  {pages} {1391} (\bibinfo {year} {2014})}\BibitemShut {NoStop}%
\bibitem [{\citenamefont {Dong}\ \emph {et~al.}(2012)\citenamefont {Dong},
  \citenamefont {Fiore}, \citenamefont {Kuzyk},\ and\ \citenamefont
  {Wang}}]{dong2012optomechanical}%
  \BibitemOpen
  \bibfield  {author} {\bibinfo {author} {\bibfnamefont {C.}~\bibnamefont
  {Dong}}, \bibinfo {author} {\bibfnamefont {V.}~\bibnamefont {Fiore}},
  \bibinfo {author} {\bibfnamefont {M.~C.}\ \bibnamefont {Kuzyk}}, \ and\
  \bibinfo {author} {\bibfnamefont {H.}~\bibnamefont {Wang}},\ }\bibfield
  {title} {\enquote {\bibinfo {title} {Optomechanical dark mode},}\ }\href@noop
  {} {\bibfield  {journal} {\bibinfo  {journal} {Science}\ }\textbf {\bibinfo
  {volume} {338}},\ \bibinfo {pages} {1609} (\bibinfo {year}
  {2012})}\BibitemShut {NoStop}%
\bibitem [{\citenamefont {Wang}\ \emph {et~al.}(2024)\citenamefont {Wang},
  \citenamefont {Zhang}, \citenamefont {Shen}, \citenamefont {Xu},
  \citenamefont {Niu}, \citenamefont {Sun}, \citenamefont {Guo},\ and\
  \citenamefont {Dong}}]{wang2024optomechanical}%
  \BibitemOpen
  \bibfield  {author} {\bibinfo {author} {\bibfnamefont {Y.}~\bibnamefont
  {Wang}}, \bibinfo {author} {\bibfnamefont {M.}~\bibnamefont {Zhang}},
  \bibinfo {author} {\bibfnamefont {Z.}~\bibnamefont {Shen}}, \bibinfo {author}
  {\bibfnamefont {G.-T.}\ \bibnamefont {Xu}}, \bibinfo {author} {\bibfnamefont
  {R.}~\bibnamefont {Niu}}, \bibinfo {author} {\bibfnamefont {F.-W.}\
  \bibnamefont {Sun}}, \bibinfo {author} {\bibfnamefont {G.-C.}\ \bibnamefont
  {Guo}}, \ and\ \bibinfo {author} {\bibfnamefont {C.-H.}\ \bibnamefont
  {Dong}},\ }\bibfield  {title} {\enquote {\bibinfo {title} {Optomechanical
  frequency comb based on multiple nonlinear dynamics},}\ }\href@noop {}
  {\bibfield  {journal} {\bibinfo  {journal} {Physical Review Letters}\
  }\textbf {\bibinfo {volume} {132}},\ \bibinfo {pages} {163603} (\bibinfo
  {year} {2024})}\BibitemShut {NoStop}%
\bibitem [{\citenamefont {Chan}\ \emph {et~al.}(2011)\citenamefont {Chan},
  \citenamefont {Alegre}, \citenamefont {Safavi-Naeini}, \citenamefont {Hill},
  \citenamefont {Krause}, \citenamefont {Gr{\"o}blacher}, \citenamefont
  {Aspelmeyer},\ and\ \citenamefont {Painter}}]{chan2011laser}%
  \BibitemOpen
  \bibfield  {author} {\bibinfo {author} {\bibfnamefont {J.}~\bibnamefont
  {Chan}}, \bibinfo {author} {\bibfnamefont {T.~M.}\ \bibnamefont {Alegre}},
  \bibinfo {author} {\bibfnamefont {A.~H.}\ \bibnamefont {Safavi-Naeini}},
  \bibinfo {author} {\bibfnamefont {J.~T.}\ \bibnamefont {Hill}}, \bibinfo
  {author} {\bibfnamefont {A.}~\bibnamefont {Krause}}, \bibinfo {author}
  {\bibfnamefont {S.}~\bibnamefont {Gr{\"o}blacher}}, \bibinfo {author}
  {\bibfnamefont {M.}~\bibnamefont {Aspelmeyer}}, \ and\ \bibinfo {author}
  {\bibfnamefont {O.}~\bibnamefont {Painter}},\ }\bibfield  {title} {\enquote
  {\bibinfo {title} {Laser cooling of a nanomechanical oscillator into its
  quantum ground state},}\ }\href@noop {} {\bibfield  {journal} {\bibinfo
  {journal} {Nature}\ }\textbf {\bibinfo {volume} {478}},\ \bibinfo {pages}
  {89} (\bibinfo {year} {2011})}\BibitemShut {NoStop}%
\bibitem [{\citenamefont {Chu}\ \emph {et~al.}(2018)\citenamefont {Chu},
  \citenamefont {Kharel}, \citenamefont {Yoon}, \citenamefont {Frunzio},
  \citenamefont {Rakich},\ and\ \citenamefont {Schoelkopf}}]{chu2018creation}%
  \BibitemOpen
  \bibfield  {author} {\bibinfo {author} {\bibfnamefont {Y.}~\bibnamefont
  {Chu}}, \bibinfo {author} {\bibfnamefont {P.}~\bibnamefont {Kharel}},
  \bibinfo {author} {\bibfnamefont {T.}~\bibnamefont {Yoon}}, \bibinfo {author}
  {\bibfnamefont {L.}~\bibnamefont {Frunzio}}, \bibinfo {author} {\bibfnamefont
  {P.~T.}\ \bibnamefont {Rakich}}, \ and\ \bibinfo {author} {\bibfnamefont
  {R.~J.}\ \bibnamefont {Schoelkopf}},\ }\bibfield  {title} {\enquote {\bibinfo
  {title} {Creation and control of multi-phonon fock states in a bulk
  acoustic-wave resonator},}\ }\href@noop {} {\bibfield  {journal} {\bibinfo
  {journal} {Nature}\ }\textbf {\bibinfo {volume} {563}},\ \bibinfo {pages}
  {666} (\bibinfo {year} {2018})}\BibitemShut {NoStop}%
\bibitem [{\citenamefont {Riedinger}\ \emph {et~al.}(2018)\citenamefont
  {Riedinger}, \citenamefont {Wallucks}, \citenamefont {Marinkovi{\'c}},
  \citenamefont {L{\"o}schnauer}, \citenamefont {Aspelmeyer}, \citenamefont
  {Hong},\ and\ \citenamefont {Gr{\"o}blacher}}]{riedinger2018remote}%
  \BibitemOpen
  \bibfield  {author} {\bibinfo {author} {\bibfnamefont {R.}~\bibnamefont
  {Riedinger}}, \bibinfo {author} {\bibfnamefont {A.}~\bibnamefont {Wallucks}},
  \bibinfo {author} {\bibfnamefont {I.}~\bibnamefont {Marinkovi{\'c}}},
  \bibinfo {author} {\bibfnamefont {C.}~\bibnamefont {L{\"o}schnauer}},
  \bibinfo {author} {\bibfnamefont {M.}~\bibnamefont {Aspelmeyer}}, \bibinfo
  {author} {\bibfnamefont {S.}~\bibnamefont {Hong}}, \ and\ \bibinfo {author}
  {\bibfnamefont {S.}~\bibnamefont {Gr{\"o}blacher}},\ }\bibfield  {title}
  {\enquote {\bibinfo {title} {Remote quantum entanglement between two
  micromechanical oscillators},}\ }\href@noop {} {\bibfield  {journal}
  {\bibinfo  {journal} {Nature}\ }\textbf {\bibinfo {volume} {556}},\ \bibinfo
  {pages} {473} (\bibinfo {year} {2018})}\BibitemShut {NoStop}%
\bibitem [{\citenamefont {Wollack}\ \emph {et~al.}(2022)\citenamefont
  {Wollack}, \citenamefont {Cleland}, \citenamefont {Gruenke}, \citenamefont
  {Wang}, \citenamefont {Arrangoiz-Arriola},\ and\ \citenamefont
  {Safavi-Naeini}}]{wollack2022quantum}%
  \BibitemOpen
  \bibfield  {author} {\bibinfo {author} {\bibfnamefont {E.~A.}\ \bibnamefont
  {Wollack}}, \bibinfo {author} {\bibfnamefont {A.~Y.}\ \bibnamefont
  {Cleland}}, \bibinfo {author} {\bibfnamefont {R.~G.}\ \bibnamefont
  {Gruenke}}, \bibinfo {author} {\bibfnamefont {Z.}~\bibnamefont {Wang}},
  \bibinfo {author} {\bibfnamefont {P.}~\bibnamefont {Arrangoiz-Arriola}}, \
  and\ \bibinfo {author} {\bibfnamefont {A.~H.}\ \bibnamefont
  {Safavi-Naeini}},\ }\bibfield  {title} {\enquote {\bibinfo {title} {Quantum
  state preparation and tomography of entangled mechanical resonators},}\
  }\href@noop {} {\bibfield  {journal} {\bibinfo  {journal} {Nature}\ }\textbf
  {\bibinfo {volume} {604}},\ \bibinfo {pages} {463} (\bibinfo {year}
  {2022})}\BibitemShut {NoStop}%
\bibitem [{\citenamefont {Qiao}\ \emph {et~al.}(2023)\citenamefont {Qiao},
  \citenamefont {Dumur}, \citenamefont {Andersson}, \citenamefont {Yan},
  \citenamefont {Chou}, \citenamefont {Grebel}, \citenamefont {Conner},
  \citenamefont {Joshi}, \citenamefont {Miller}, \citenamefont {Povey} \emph
  {et~al.}}]{qiao2023splitting}%
  \BibitemOpen
  \bibfield  {author} {\bibinfo {author} {\bibfnamefont {H.}~\bibnamefont
  {Qiao}}, \bibinfo {author} {\bibfnamefont {{\'E}.}~\bibnamefont {Dumur}},
  \bibinfo {author} {\bibfnamefont {G.}~\bibnamefont {Andersson}}, \bibinfo
  {author} {\bibfnamefont {H.}~\bibnamefont {Yan}}, \bibinfo {author}
  {\bibfnamefont {M.-H.}\ \bibnamefont {Chou}}, \bibinfo {author}
  {\bibfnamefont {J.}~\bibnamefont {Grebel}}, \bibinfo {author} {\bibfnamefont
  {C.}~\bibnamefont {Conner}}, \bibinfo {author} {\bibfnamefont
  {Y.}~\bibnamefont {Joshi}}, \bibinfo {author} {\bibfnamefont
  {J.}~\bibnamefont {Miller}}, \bibinfo {author} {\bibfnamefont
  {R.}~\bibnamefont {Povey}},  \emph {et~al.},\ }\bibfield  {title} {\enquote
  {\bibinfo {title} {Splitting phonons: Building a platform for linear
  mechanical quantum computing},}\ }\href@noop {} {\bibfield  {journal}
  {\bibinfo  {journal} {Science}\ }\textbf {\bibinfo {volume} {380}},\ \bibinfo
  {pages} {1030} (\bibinfo {year} {2023})}\BibitemShut {NoStop}%
\bibitem [{\citenamefont {Yang}\ \emph {et~al.}(2024)\citenamefont {Yang},
  \citenamefont {Kladari{\'c}}, \citenamefont {Drimmer}, \citenamefont {von
  L{\"u}pke}, \citenamefont {Lenterman}, \citenamefont {Bus}, \citenamefont
  {Marti}, \citenamefont {Fadel},\ and\ \citenamefont
  {Chu}}]{yang2024mechanical}%
  \BibitemOpen
  \bibfield  {author} {\bibinfo {author} {\bibfnamefont {Y.}~\bibnamefont
  {Yang}}, \bibinfo {author} {\bibfnamefont {I.}~\bibnamefont {Kladari{\'c}}},
  \bibinfo {author} {\bibfnamefont {M.}~\bibnamefont {Drimmer}}, \bibinfo
  {author} {\bibfnamefont {U.}~\bibnamefont {von L{\"u}pke}}, \bibinfo {author}
  {\bibfnamefont {D.}~\bibnamefont {Lenterman}}, \bibinfo {author}
  {\bibfnamefont {J.}~\bibnamefont {Bus}}, \bibinfo {author} {\bibfnamefont
  {S.}~\bibnamefont {Marti}}, \bibinfo {author} {\bibfnamefont
  {M.}~\bibnamefont {Fadel}}, \ and\ \bibinfo {author} {\bibfnamefont
  {Y.}~\bibnamefont {Chu}},\ }\bibfield  {title} {\enquote {\bibinfo {title} {A
  mechanical qubit},}\ }\href@noop {} {\bibfield  {journal} {\bibinfo
  {journal} {Science}\ }\textbf {\bibinfo {volume} {386}},\ \bibinfo {pages}
  {783} (\bibinfo {year} {2024})}\BibitemShut {NoStop}%
\bibitem [{\citenamefont {Zivari}\ \emph {et~al.}(2023)\citenamefont {Zivari},
  \citenamefont {Fiaschi}, \citenamefont {Scarpelli}, \citenamefont {Jansen},
  \citenamefont {Burgwal}, \citenamefont {Verhagen},\ and\ \citenamefont
  {Gr{\"o}blacher}}]{zivari2023single}%
  \BibitemOpen
  \bibfield  {author} {\bibinfo {author} {\bibfnamefont {A.}~\bibnamefont
  {Zivari}}, \bibinfo {author} {\bibfnamefont {N.}~\bibnamefont {Fiaschi}},
  \bibinfo {author} {\bibfnamefont {L.}~\bibnamefont {Scarpelli}}, \bibinfo
  {author} {\bibfnamefont {M.}~\bibnamefont {Jansen}}, \bibinfo {author}
  {\bibfnamefont {R.}~\bibnamefont {Burgwal}}, \bibinfo {author} {\bibfnamefont
  {E.}~\bibnamefont {Verhagen}}, \ and\ \bibinfo {author} {\bibfnamefont
  {S.}~\bibnamefont {Gr{\"o}blacher}},\ }\bibfield  {title} {\enquote {\bibinfo
  {title} {A single-phonon directional coupler},}\ }\href@noop {} {\bibfield
  {journal} {\bibinfo  {journal} {arXiv preprint arXiv:2312.04414}\ } (\bibinfo
  {year} {2023})}\BibitemShut {NoStop}%
\bibitem [{\citenamefont {Arrangoiz-Arriola}\ \emph {et~al.}(2019)\citenamefont
  {Arrangoiz-Arriola}, \citenamefont {Wollack}, \citenamefont {Wang},
  \citenamefont {Pechal}, \citenamefont {Jiang}, \citenamefont {McKenna},
  \citenamefont {Witmer}, \citenamefont {Van~Laer},\ and\ \citenamefont
  {Safavi-Naeini}}]{arrangoiz2019resolving}%
  \BibitemOpen
  \bibfield  {author} {\bibinfo {author} {\bibfnamefont {P.}~\bibnamefont
  {Arrangoiz-Arriola}}, \bibinfo {author} {\bibfnamefont {E.~A.}\ \bibnamefont
  {Wollack}}, \bibinfo {author} {\bibfnamefont {Z.}~\bibnamefont {Wang}},
  \bibinfo {author} {\bibfnamefont {M.}~\bibnamefont {Pechal}}, \bibinfo
  {author} {\bibfnamefont {W.}~\bibnamefont {Jiang}}, \bibinfo {author}
  {\bibfnamefont {T.~P.}\ \bibnamefont {McKenna}}, \bibinfo {author}
  {\bibfnamefont {J.~D.}\ \bibnamefont {Witmer}}, \bibinfo {author}
  {\bibfnamefont {R.}~\bibnamefont {Van~Laer}}, \ and\ \bibinfo {author}
  {\bibfnamefont {A.~H.}\ \bibnamefont {Safavi-Naeini}},\ }\bibfield  {title}
  {\enquote {\bibinfo {title} {Resolving the energy levels of a nanomechanical
  oscillator},}\ }\href@noop {} {\bibfield  {journal} {\bibinfo  {journal}
  {Nature}\ }\textbf {\bibinfo {volume} {571}},\ \bibinfo {pages} {537}
  (\bibinfo {year} {2019})}\BibitemShut {NoStop}%
\bibitem [{\citenamefont {Kuang}\ \emph {et~al.}(2023)\citenamefont {Kuang},
  \citenamefont {Huang}, \citenamefont {Xiong}, \citenamefont {Zuo},
  \citenamefont {Han}, \citenamefont {Nori}, \citenamefont {Qiu}, \citenamefont
  {Luo}, \citenamefont {Jing},\ and\ \citenamefont
  {Xiao}}]{kuang2023nonlinear}%
  \BibitemOpen
  \bibfield  {author} {\bibinfo {author} {\bibfnamefont {T.}~\bibnamefont
  {Kuang}}, \bibinfo {author} {\bibfnamefont {R.}~\bibnamefont {Huang}},
  \bibinfo {author} {\bibfnamefont {W.}~\bibnamefont {Xiong}}, \bibinfo
  {author} {\bibfnamefont {Y.}~\bibnamefont {Zuo}}, \bibinfo {author}
  {\bibfnamefont {X.}~\bibnamefont {Han}}, \bibinfo {author} {\bibfnamefont
  {F.}~\bibnamefont {Nori}}, \bibinfo {author} {\bibfnamefont {C.-W.}\
  \bibnamefont {Qiu}}, \bibinfo {author} {\bibfnamefont {H.}~\bibnamefont
  {Luo}}, \bibinfo {author} {\bibfnamefont {H.}~\bibnamefont {Jing}}, \ and\
  \bibinfo {author} {\bibfnamefont {G.}~\bibnamefont {Xiao}},\ }\bibfield
  {title} {\enquote {\bibinfo {title} {Nonlinear multi-frequency phonon lasers
  with active levitated optomechanics},}\ }\href@noop {} {\bibfield  {journal}
  {\bibinfo  {journal} {Nature Physics}\ }\textbf {\bibinfo {volume} {19}},\
  \bibinfo {pages} {414} (\bibinfo {year} {2023})}\BibitemShut {NoStop}%
\bibitem [{\citenamefont {Silverstone}\ \emph {et~al.}(2014)\citenamefont
  {Silverstone}, \citenamefont {Bonneau}, \citenamefont {Ohira}, \citenamefont
  {Suzuki}, \citenamefont {Yoshida}, \citenamefont {Iizuka}, \citenamefont
  {Ezaki}, \citenamefont {Natarajan}, \citenamefont {Tanner}, \citenamefont
  {Hadfield} \emph {et~al.}}]{silverstone2014chip}%
  \BibitemOpen
  \bibfield  {author} {\bibinfo {author} {\bibfnamefont {J.~W.}\ \bibnamefont
  {Silverstone}}, \bibinfo {author} {\bibfnamefont {D.}~\bibnamefont
  {Bonneau}}, \bibinfo {author} {\bibfnamefont {K.}~\bibnamefont {Ohira}},
  \bibinfo {author} {\bibfnamefont {N.}~\bibnamefont {Suzuki}}, \bibinfo
  {author} {\bibfnamefont {H.}~\bibnamefont {Yoshida}}, \bibinfo {author}
  {\bibfnamefont {N.}~\bibnamefont {Iizuka}}, \bibinfo {author} {\bibfnamefont
  {M.}~\bibnamefont {Ezaki}}, \bibinfo {author} {\bibfnamefont {C.~M.}\
  \bibnamefont {Natarajan}}, \bibinfo {author} {\bibfnamefont {M.~G.}\
  \bibnamefont {Tanner}}, \bibinfo {author} {\bibfnamefont {R.~H.}\
  \bibnamefont {Hadfield}},  \emph {et~al.},\ }\bibfield  {title} {\enquote
  {\bibinfo {title} {On-chip quantum interference between silicon photon-pair
  sources},}\ }\href@noop {} {\bibfield  {journal} {\bibinfo  {journal} {Nature
  Photonics}\ }\textbf {\bibinfo {volume} {8}},\ \bibinfo {pages} {104}
  (\bibinfo {year} {2014})}\BibitemShut {NoStop}%
\bibitem [{\citenamefont {Afek}\ \emph {et~al.}(2010)\citenamefont {Afek},
  \citenamefont {Ambar},\ and\ \citenamefont {Silberberg}}]{afek2010high}%
  \BibitemOpen
  \bibfield  {author} {\bibinfo {author} {\bibfnamefont {I.}~\bibnamefont
  {Afek}}, \bibinfo {author} {\bibfnamefont {O.}~\bibnamefont {Ambar}}, \ and\
  \bibinfo {author} {\bibfnamefont {Y.}~\bibnamefont {Silberberg}},\ }\bibfield
   {title} {\enquote {\bibinfo {title} {High-noon states by mixing quantum and
  classical light},}\ }\href@noop {} {\bibfield  {journal} {\bibinfo  {journal}
  {Science}\ }\textbf {\bibinfo {volume} {328}},\ \bibinfo {pages} {879}
  (\bibinfo {year} {2010})}\BibitemShut {NoStop}%
\bibitem [{\citenamefont {Wollman}\ \emph {et~al.}(2015)\citenamefont
  {Wollman}, \citenamefont {Lei}, \citenamefont {Weinstein}, \citenamefont
  {Suh}, \citenamefont {Kronwald}, \citenamefont {Marquardt}, \citenamefont
  {Clerk},\ and\ \citenamefont {Schwab}}]{wollman2015quantum}%
  \BibitemOpen
  \bibfield  {author} {\bibinfo {author} {\bibfnamefont {E.~E.}\ \bibnamefont
  {Wollman}}, \bibinfo {author} {\bibfnamefont {C.}~\bibnamefont {Lei}},
  \bibinfo {author} {\bibfnamefont {A.}~\bibnamefont {Weinstein}}, \bibinfo
  {author} {\bibfnamefont {J.}~\bibnamefont {Suh}}, \bibinfo {author}
  {\bibfnamefont {A.}~\bibnamefont {Kronwald}}, \bibinfo {author}
  {\bibfnamefont {F.}~\bibnamefont {Marquardt}}, \bibinfo {author}
  {\bibfnamefont {A.~A.}\ \bibnamefont {Clerk}}, \ and\ \bibinfo {author}
  {\bibfnamefont {K.}~\bibnamefont {Schwab}},\ }\bibfield  {title} {\enquote
  {\bibinfo {title} {Quantum squeezing of motion in a mechanical resonator},}\
  }\href@noop {} {\bibfield  {journal} {\bibinfo  {journal} {Science}\ }\textbf
  {\bibinfo {volume} {349}},\ \bibinfo {pages} {952} (\bibinfo {year}
  {2015})}\BibitemShut {NoStop}%
\bibitem [{\citenamefont {Chan}\ \emph {et~al.}(2012)\citenamefont {Chan},
  \citenamefont {Safavi-Naeini}, \citenamefont {Hill}, \citenamefont
  {Meenehan},\ and\ \citenamefont {Painter}}]{chan2012optimized}%
  \BibitemOpen
  \bibfield  {author} {\bibinfo {author} {\bibfnamefont {J.}~\bibnamefont
  {Chan}}, \bibinfo {author} {\bibfnamefont {A.~H.}\ \bibnamefont
  {Safavi-Naeini}}, \bibinfo {author} {\bibfnamefont {J.~T.}\ \bibnamefont
  {Hill}}, \bibinfo {author} {\bibfnamefont {S.}~\bibnamefont {Meenehan}}, \
  and\ \bibinfo {author} {\bibfnamefont {O.}~\bibnamefont {Painter}},\
  }\bibfield  {title} {\enquote {\bibinfo {title} {Optimized optomechanical
  crystal cavity with acoustic radiation shield},}\ }\href@noop {} {\bibfield
  {journal} {\bibinfo  {journal} {Applied Physics Letters}\ }\textbf {\bibinfo
  {volume} {101}} (\bibinfo {year} {2012})}\BibitemShut {NoStop}%
\bibitem [{\citenamefont {Wang}\ \emph {et~al.}(2023)\citenamefont {Wang},
  \citenamefont {Shi}, \citenamefont {Kuang}, \citenamefont {Xi}, \citenamefont
  {Wan}, \citenamefont {Shen}, \citenamefont {Wang}, \citenamefont {Xu},
  \citenamefont {Sun}, \citenamefont {Zou} \emph
  {et~al.}}]{wang2023realization}%
  \BibitemOpen
  \bibfield  {author} {\bibinfo {author} {\bibfnamefont {Y.}~\bibnamefont
  {Wang}}, \bibinfo {author} {\bibfnamefont {Z.-P.}\ \bibnamefont {Shi}},
  \bibinfo {author} {\bibfnamefont {H.-Y.}\ \bibnamefont {Kuang}}, \bibinfo
  {author} {\bibfnamefont {X.}~\bibnamefont {Xi}}, \bibinfo {author}
  {\bibfnamefont {S.}~\bibnamefont {Wan}}, \bibinfo {author} {\bibfnamefont
  {Z.}~\bibnamefont {Shen}}, \bibinfo {author} {\bibfnamefont {P.-Y.}\
  \bibnamefont {Wang}}, \bibinfo {author} {\bibfnamefont {G.-T.}\ \bibnamefont
  {Xu}}, \bibinfo {author} {\bibfnamefont {X.}~\bibnamefont {Sun}}, \bibinfo
  {author} {\bibfnamefont {C.-L.}\ \bibnamefont {Zou}},  \emph {et~al.},\
  }\bibfield  {title} {\enquote {\bibinfo {title} {Realization of quantum
  ground state in an optomechanical crystal cavity},}\ }\href@noop {}
  {\bibfield  {journal} {\bibinfo  {journal} {Science China Physics, Mechanics
  \& Astronomy}\ }\textbf {\bibinfo {volume} {66}},\ \bibinfo {pages} {124213}
  (\bibinfo {year} {2023})}\BibitemShut {NoStop}%
\bibitem [{\citenamefont {Kuzmich}\ \emph {et~al.}(2003)\citenamefont
  {Kuzmich}, \citenamefont {Bowen}, \citenamefont {Boozer}, \citenamefont
  {Boca}, \citenamefont {Chou}, \citenamefont {Duan},\ and\ \citenamefont
  {Kimble}}]{kuzmich2003generation}%
  \BibitemOpen
  \bibfield  {author} {\bibinfo {author} {\bibfnamefont {A.}~\bibnamefont
  {Kuzmich}}, \bibinfo {author} {\bibfnamefont {W.}~\bibnamefont {Bowen}},
  \bibinfo {author} {\bibfnamefont {A.}~\bibnamefont {Boozer}}, \bibinfo
  {author} {\bibfnamefont {A.}~\bibnamefont {Boca}}, \bibinfo {author}
  {\bibfnamefont {C.}~\bibnamefont {Chou}}, \bibinfo {author} {\bibfnamefont
  {L.-M.}\ \bibnamefont {Duan}}, \ and\ \bibinfo {author} {\bibfnamefont
  {H.}~\bibnamefont {Kimble}},\ }\bibfield  {title} {\enquote {\bibinfo {title}
  {Generation of nonclassical photon pairs for scalable quantum communication
  with atomic ensembles},}\ }\href@noop {} {\bibfield  {journal} {\bibinfo
  {journal} {Nature}\ }\textbf {\bibinfo {volume} {423}},\ \bibinfo {pages}
  {731} (\bibinfo {year} {2003})}\BibitemShut {NoStop}%
\bibitem [{\citenamefont {Lei}\ \emph {et~al.}(2016)\citenamefont {Lei},
  \citenamefont {Weinstein}, \citenamefont {Suh}, \citenamefont {Wollman},
  \citenamefont {Kronwald}, \citenamefont {Marquardt}, \citenamefont {Clerk},\
  and\ \citenamefont {Schwab}}]{lei2016quantum}%
  \BibitemOpen
  \bibfield  {author} {\bibinfo {author} {\bibfnamefont {C.}~\bibnamefont
  {Lei}}, \bibinfo {author} {\bibfnamefont {A.}~\bibnamefont {Weinstein}},
  \bibinfo {author} {\bibfnamefont {J.}~\bibnamefont {Suh}}, \bibinfo {author}
  {\bibfnamefont {E.}~\bibnamefont {Wollman}}, \bibinfo {author} {\bibfnamefont
  {A.}~\bibnamefont {Kronwald}}, \bibinfo {author} {\bibfnamefont
  {F.}~\bibnamefont {Marquardt}}, \bibinfo {author} {\bibfnamefont
  {A.}~\bibnamefont {Clerk}}, \ and\ \bibinfo {author} {\bibfnamefont
  {K.}~\bibnamefont {Schwab}},\ }\bibfield  {title} {\enquote {\bibinfo {title}
  {Quantum nondemolition measurement of a quantum squeezed state beyond the 3
  db limit},}\ }\href@noop {} {\bibfield  {journal} {\bibinfo  {journal}
  {Physical review letters}\ }\textbf {\bibinfo {volume} {117}},\ \bibinfo
  {pages} {100801} (\bibinfo {year} {2016})}\BibitemShut {NoStop}%
\bibitem [{\citenamefont {Chen}\ \emph {et~al.}(2024)\citenamefont {Chen},
  \citenamefont {Korsch}, \citenamefont {Kersul}, \citenamefont {Benevides},
  \citenamefont {Yu}, \citenamefont {Alegre},\ and\ \citenamefont
  {Gr{\"o}blacher}}]{chen2024bandwidth}%
  \BibitemOpen
  \bibfield  {author} {\bibinfo {author} {\bibfnamefont {L.}~\bibnamefont
  {Chen}}, \bibinfo {author} {\bibfnamefont {A.~R.}\ \bibnamefont {Korsch}},
  \bibinfo {author} {\bibfnamefont {C.~M.}\ \bibnamefont {Kersul}}, \bibinfo
  {author} {\bibfnamefont {R.}~\bibnamefont {Benevides}}, \bibinfo {author}
  {\bibfnamefont {Y.}~\bibnamefont {Yu}}, \bibinfo {author} {\bibfnamefont
  {T.~P.~M.}\ \bibnamefont {Alegre}}, \ and\ \bibinfo {author} {\bibfnamefont
  {S.}~\bibnamefont {Gr{\"o}blacher}},\ }\bibfield  {title} {\enquote {\bibinfo
  {title} {Bandwidth-tunable telecom single photons enabled by low-noise
  optomechanical transduction},}\ }\href@noop {} {\bibfield  {journal}
  {\bibinfo  {journal} {arXiv preprint arXiv:2410.10947}\ } (\bibinfo {year}
  {2024})}\BibitemShut {NoStop}%

\end{thebibliography}

\emph{Acknowledgments.-}The authors would like to thank H. Wang and
Y. Guo for discussions. This work was supported by the Innovation
program for Quantum Science and Technology (2021ZD0303203), National
Natural Science Foundation of China (Grant No.12293052, 12293050,
11934012, 12474394, 12104442, 12447139 and 92250302), the China Postdoctoral
Science Foundation (Grant No. 2024T008AH), the Natural Science Foundation
of Anhui Province (Grant No. 2308085J12 and 2408085QA021), the Fundamental
Research Funds for the Central Universities, USTC Major Frontier Research
Program (LS2030000002), and the CAS Project for Young Scientists in
Basic Research (YSBR-069). This work was partially carried out at
the USTC Center for Micro and Nanoscale Research and Fabrication.

\emph{Author contributions.-}Y.W., Z.S. and C.H.D. conceived the experiment.
Y.W., H.Y.K., and S.W. performed the device design and fabrication.
Y.W., Z.S., Z.P.S.and C.H.D. performed the measurements, analyzed
the data. M.Z. and Y.W. performed the theoretical simulation and classical-quantum
criterion derivation, F.W.S. provided theoretical support. Y.W., Z.S.,
M.Z. and C.H.D. wrote the manuscript. C.H.D. and G.C.G. supervised
the project.

\emph{Competing interests.-}The authors declare no competing interests.

\emph{Additional information.-}Supplementary information is available
online. Correspondence and requests for materials should be addressed
to C.-H.D.
\end{document}